\documentclass{JHEP3}
\usepackage{epsfig,multicol}
\def\be{\begin{equation}}
\def\ee{\end{equation}}    
\def\baray{\begin{eqnarray}}
\def\earay{\end{eqnarray}}
\def\dper{d_\perp}
\def\rp{r_\perp}
\def\rpl{r_\parallel}
\def\Vp{V_\perp}
\def\Vpl{V_\parallel}
\def\lp{\ell_\perp}  
\def\lpl{\ell_\parallel}

\def\vpl{v_\parallel}
\def\a{{\alpha^\prime}}     
\newcommand\Tr[1]{\mathbb{\mathrm{tr}}_{#1}}

\title{Brane Interaction as the Origin of Inflation}
\author{Nicholas Jones, Horace Stoica and S.-H. Henry Tye \\
	Laboratory of Elementary Particle Physics, Cornell University, 
Ithaca, NY 14853\\
	E-mail: \email{nick.jones@cornell.edu},  
	\email{fhs3@mail.lns.cornell.edu},
	\email{tye@mail.lns.cornell.edu}}
\abstract{
We reanalyze brane inflation with brane-brane interactions 
at an angle, which include the special case of brane-anti-brane 
interaction. If nature is described by a stringy realization
of the brane world scenario today (with arbitrary compactification),
and if some additional branes were present in the early universe, 
we find that an inflationary epoch is generically quite natural, 
ending with a big bang when the last branes collide. 
In an interesting brane inflationary scenario suggested by
generic string model-building, we use the density perturbation observed 
in the cosmic microwave background and the coupling unification to 
find that the string scale is comparable to the $GUT$ scale.}

\preprint{}

\begin{document}

\section{Introduction}

By now, the inflationary universe \cite{guth}
is generally recognized to be the most likely scenario that
explains the origin of the big bang. So far, its predictions of
the flatness of the universe and the almost scale-invariant
power spectrum of the density perturbation that seeds structure
formations are in very good agreement with the cosmic microwave
background (CMB) observations \cite{cobe,new}.
In standard inflationary models \cite{revs}, the physics lies in
the inflaton potential. However, conventional physics (such as
the standard electroweak and strong interaction model or grand 
unified theories) does not yield an inflaton potential that
agrees with observations (such as enough number of e-foldings,
the amplitude of the density perturbation etc.). In the past
two decades, numerous phenomenological potentials that give the
correct inflationary properties have been proposed, most with
little input from fundamental physics.
In general, a potential that yields enough e-foldings and the 
correct magnitude of density perturbation requires some fine-tuning.
However, such fine-tuning is in general not preserved by quantum
corrections. Recently, the brane inflationary
scenario \cite{dvali-tye} was proposed, where
the inflaton is identified with an inter-brane separation, while
the inflaton potential emerged from brane interactions 
that is well-studied in string theory \cite{Polchinski1}.
In particular, exchanges of closed string (bulk)
modes between branes (the lowest order of which can be determined 
by the one-loop partition function of the open string spectrum)
determine the form of the inflaton potential
during the slow-roll epoch. 
Inflation ends when the branes
collide, heating the universe that starts the big bang.
This visualization of the brane dynamics
allows one to implement inflation physics pictorially.  

The brane inflation scenario may be realized in a variety of ways 
\cite{dvali,burgess,rabadan,collection}. 
One possibility is the brane-anti-brane interaction
and annihilation \cite{dvali,burgess}. Although this potential is 
too steep for inflation, a particularly appealing scenario 
proposed by Burgess etc.~\cite{burgess} shows that a hypercubic 
compactification of extra dimensions flattens the potential to yield 
enough e-foldings for a viable inflationary epoch. 
The brane inflationary scenario for branes at a fixed angle is studied
by Garcia-Bellido etc. \cite{rabadan}, which is a generalization of the 
brane-anti-brane scenario and is quite natural in 
string-model building. We re-analyze the brane inflationary 
scenario by combining and generalizing
these two observations and find that the result is very robust.
For cases where inflation is generic, we use the CMB data to determine 
the string scale  to be $M_s \lesssim M_{GUT}$. 

From experimental data and assuming MSSM, the gauge couplings of 
the standard model unify at the $GUT$ scale \cite{gut}:
\be
\label{gutvalue}
\alpha(GUT) \simeq 1/25  \quad  \quad  M_{GUT} \simeq 2 \times 10^{16} GeV
\ee            
In the brane world scenario, the standard model particles 
are open string (brane) modes while graviton and other closed string 
modes are bulk modes. This may be naturally realized in 
Type I (or orientifold) string theories, with string scale 
$M_s= 1/\sqrt{{\alpha \prime}}$
and large compactified dimensions.
After compactification to 4-D spacetime, the Planck mass, 
$M_P = (8\pi G_N)^{-1/2} = 2.42 \times 10^{18}$ GeV, is given 
by, via dimensional reduction \cite{shiu,alwis}  
\be
\label{MP1}
g_s^2 M_P^2 = \frac{M_s^8 V}{(2 \pi)^6 \pi}=\frac{M_s^8 V_t V_P}
{(2 \pi)^6 \pi} 
\ee
Here $V= V_t V_P$ is the 6-D compactification volume, 
where the $(p-3)$ dimensions of the stack of D$p$-branes we live in 
are compactified with volume $V_t$.
Here $g_s=e^{\phi_D}$ is the string coupling determined by the 
dilaton expectation value $\phi_D$, which is related to 
the gauge coupling $\alpha$ via
\baray
\label{gsalpha}
g_s= 2 M_s^{p-3} V_t (2 \pi)^{-(p-3)} \alpha(M_s) 
= 2  v_t \alpha(M_s) 
\earay
The presence of the factor $(2 \pi)^6 \pi$ in Eq.(\ref{MP1})
appears because the 10-D Newton's constant $G_{10}$ in 
Type I string theory is given by \cite{alwis,Polchinski1}
\be
\label{g10}
8 \pi G_{10}= \kappa^2 = \frac{g_s^2 (2 \pi)^7}{2 M_s^8}
\ee  
In general, $M_{GUT} \ge M_s$. Crudely speaking, if $M_{GUT} \simeq M_s$, 
the gauge couplings are unified at the string scale via conventional 
(renormalization group) logarithmic running. If $M_{GUT} > M_s$, 
then $v_t > 1$ and the gauge couplings will go through 
a period of power running, unifying at the string scale \cite{dienes}.
However, the present experimental data does not determine the value of 
$M_s$. It can take value anywhere between 1 TeV to $M_{GUT}$ \cite{add}.

The key data in the cosmic microwave background (CMB) we use are
the density perturbation magnitude measured by COBE \cite{cobe} and its 
power spectrum index $n$ \cite{new}
\be
\label{datanew}
\delta_H = 1.9 \times 10^{-5} \quad \quad |n-1|<0.1
\ee   
Our crucial assumptions are that the cosmological 
constant is negligible after inflation and that the radions (moduli of 
the compact dimensions) and the dilaton are stabilized by some 
unknown physics. This assumption precludes the dynamics of these moduli
from playing a role in inflation. (Otherwise, the picture becomes much 
more complicated and model-dependent. We shall briefly discuss
how they may play a role in an inflationary scenario \cite{ira,radion}.)

For two branes at a distance and at angle $\theta$, the 
parameters coming from string theory are essentially $M_s$ 
and $0 < \theta \le \pi$, as well as the integers $p \ge 3$ and 
$\dper$, where $p + \dper \le 9$.
The generic inflationary picture that emerges in our analysis may be 
summarized as follows:
\begin{itemize}
\item The branes may start out wrinkled and curved, with matter 
density or defects on them and in the bulk. They may even intersect 
each other in the 
uncompactified directions. Generically, there is a density of strings
stretched between branes as well. They look like massive particles
to the 4-D observer. Branes with zero, one or two uncompactified dimensions 
are defects that look like point-like objects, cosmic strings and 
domain walls, respectively. Fortunately, inflation will smooth 
out the wrinkles, inflate away the curvature and the 
matter/defect/stretched-string densities, and red shift the intersections
exponentially far away from any generic point on the brane, so one
may consider the branes with 3 uncompactified dimensions to be 
essentially parallel in the uncompactified 
directions, with empty branes and empty bulk.

\DOUBLEFIGURE[t]{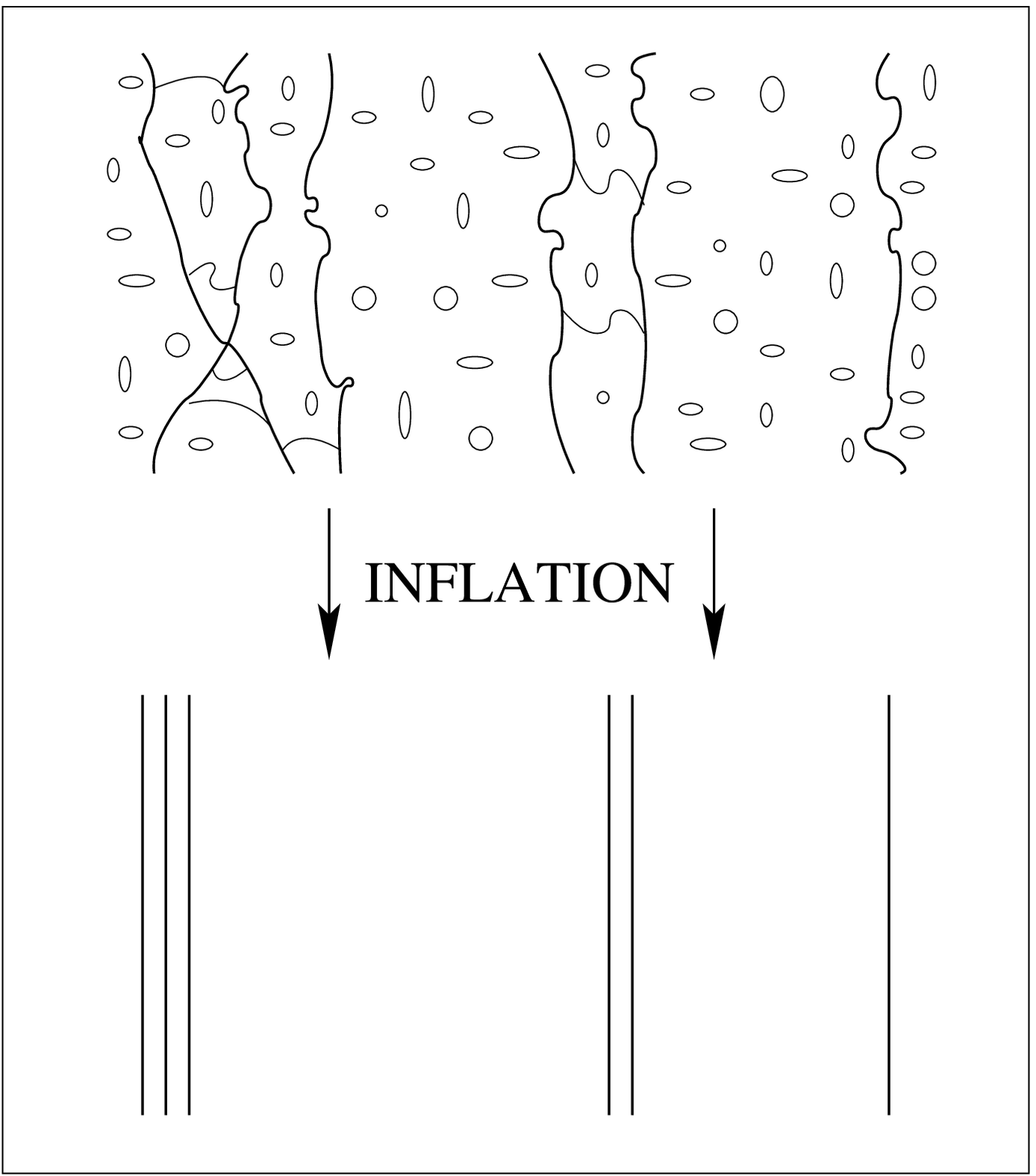, width=.4\textwidth}
	{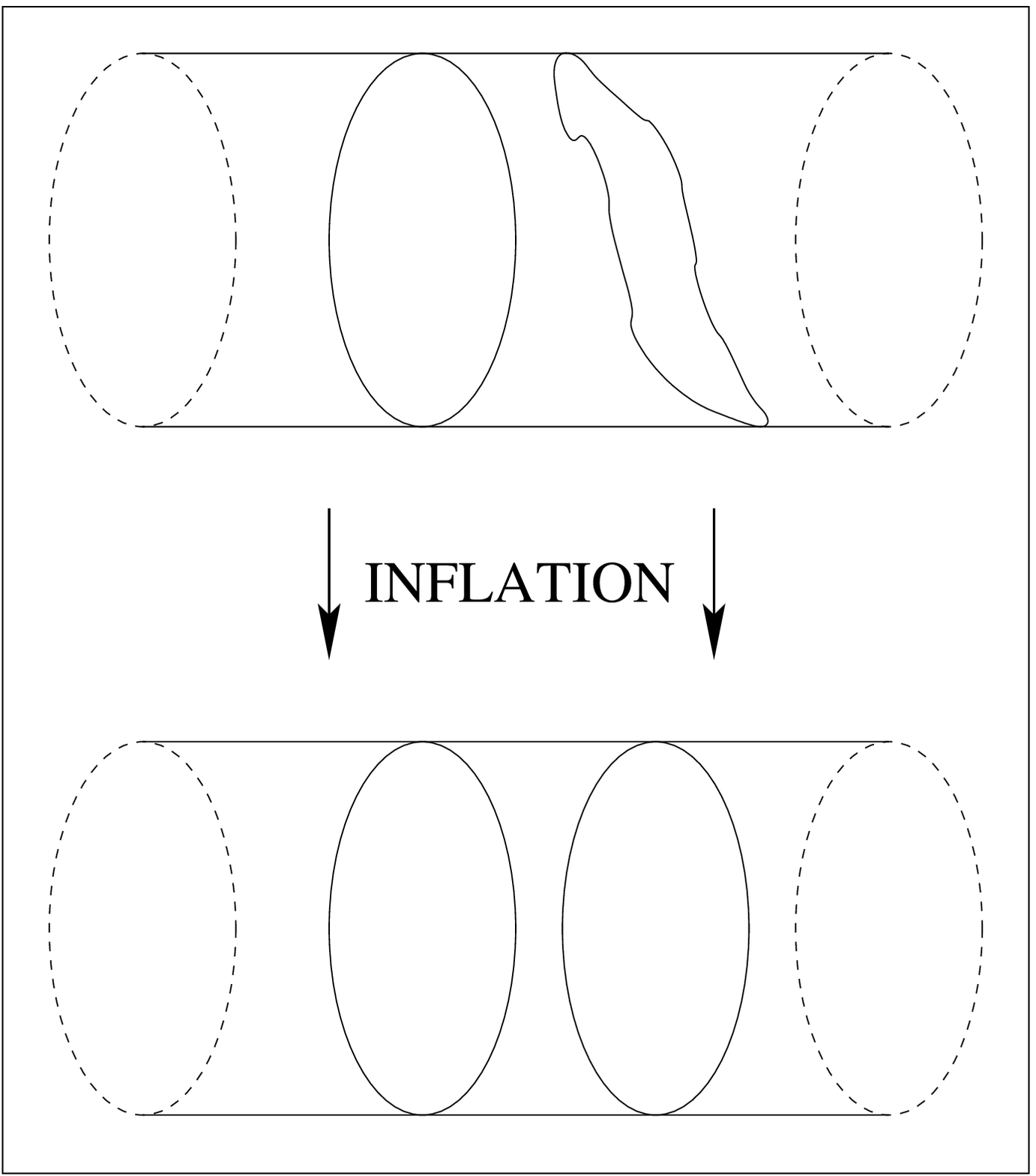, width=.4\textwidth}
	{Inflating away brane and bulk fluctuations.}
	{Inflating away non-wrapping angles between branes.}

\item Generically the inflaton potential for a brane-anti-brane system
is too steep for inflation \cite{dvali,burgess}.
Ref.\cite{burgess} shows that in a hypercubic toroidal compactification,
the images of the brane also exert forces on the anti-brane. 
For separations comparable to the compactification size, the forces 
tend to cancel. 
This softens the potential so there may be enough inflation. 
If the branes start out far enough apart, there will be a
slow-roll period for inflation before they collide.
It is sensible to ask what is the probability ${\it P}$ of sufficient 
inflation if the brane and the anti-brane are randomly placed inside the 
compactified volume. Depending on the number $\dper$ of large extra 
dimensions, we find the probability  ${\it P \sim  (3 \%)^{\dper}}$ 
This means that brane-anti-brane inflation is not very likely.
\item We find that hypercubic compactification is not necessary.
For the same brane interaction and compactification volume, we find 
that ${\it P}$ for a generic compactification is comparable to
that for a hypercubic toroidal compactification.
This point is illustrated with a rectangular torus.      
\item Ref.\cite{rabadan} argues that branes at angles improves the 
situation. A D$p$-brane differs from an anti-D$p$-brane only by an 
opposite Ramond-Ramond (R-R) charge. This charge corresponds to an 
orientation, so the special case when the angle $\theta=\pi$ corresponds 
to the brane-anti-brane case. Two branes at generic angle $\theta$ 
collide to form a lower energy brane (while brane-anti-brane annihilates). 
For two randomly placed branes at an angle, the probability ${\it P}$ 
of sufficient inflation increases to close to unity as the angle 
becomes small. Besides the factor of $(2 \pi)^6 \pi$ in Eq.(\ref{MP1})
and the inclusion of the compactification effect,
our analysis differs from Ref.\cite{rabadan} in the choice of the vacuum 
energy during inflation. Also, we believe it is more natural to choose the
rectangular torus (maybe together with a large winding number) to obtain a 
small $\theta$. In this case, ${\it P} \propto \theta^{-\dper}$.
So, for $\theta \lesssim 1/{10}$ and generic compactification,
${\it P}$ is of order unity. This implies that inflation is quite 
generic with a relatively small $\theta$.
As $\theta$ becomes smaller, the compactification lattice effect 
becomes less important. For small enough angle ({\em i.e.}, 
$\theta \lesssim 10^{-2} $), the compactification lattice effect
is negligible. In this case, ${\it P}(\theta)$ is very close to unity.
\item The magnitude of the density perturbation $\delta_H$ 
(\ref{datanew}) may be used to fix the string scale. 
Generically, we have $M_s \lesssim M_{GUT}$.
In a brane inflationary scenario suggested by
the analysis and generic orientifold construction, we find that
$M_s \sim M_{GUT}$. It will be very interesting to see how close $M_s$
is to $M_{GUT}$ in a more realistic string model.
\item Soon after inflation ends, the ground state open string modes 
become tachyonic, which leads to tachyon condensation and brane 
annihilation/collision, reheating the universe. 
This is an explicit realization of hybrid inflation \cite{Linde}.
It is well-known that $p$-brane collision generically allows the 
production of lower-dimensional $(p-2k)$-branes. 
Since the inflaton is different from the
tachyon modes that are responsible for defect production, the 
production of defects is generically a serious problem.
We argue that the dangerous productions of domain walls and 
monopole-like defects are actually absent, while the production
of the more palatable cosmic strings may happen via the Kibble mechanism. 
In fact, the production of cosmic strings can be considered as a 
prediction of this brane inflationary scenario. 
A more careful analysis is clearly needed to determine the 
cosmic string density.

\end{itemize}

Generically, bulk modes have gravitational strength couplings and so
brane modes are more likely inflaton candidates. In string theory, 
vacuum expectation values of such scalar brane modes are simply 
brane separations, which are the inflatons in brane inflation.
Compactification must take place and it always softens the inflaton
potential to improve ${\it P}$ for inflation. Branes at angles is more 
generic than a brane-anti-brane pair, and 
a small $\theta$ is easily realized if two compactification sizes
have a ratio $\sim \theta$. These lead us to argue that brane 
inflation is probably quite generic. 
In this sense, brane interaction may be considered as
an explanation of the origin of inflation.

The attractive force between two branes is stronger for smaller 
separation and larger angle $\theta$.
If the early universe starts with a number of branes (as will 
be the case if the universe starts out to be very hot, or if there is 
a higher-dimensional brane-anti-brane pair or a non-BPS brane),
probably the universe goes through multiple inflationary epochs,
with the last collision involving two branes that start out relatively 
far apart and at small angle. The inflation we have been discussing 
is the epoch just before their collision. It is interesting to 
study this scenario in greater detail.

In Section 2, we write down the effective theory for multi-branes.
In Section 3, we set up the brane world scenario just before the final
inflation takes place, for a brane-anti-brane pair and for two branes 
at a fixed angle. 
The application of this brane pair to inflation is discussed in 
Section 4, where we find that two branes at a small angle in a 
compactified volume generically generates inflation. We also discuss 
the issues of reheating and possible defect productions after inflation.
In Section 5, we commend on the bulk modes and the robustness of 
$M_s \lesssim M{GUT}$.
The summary and remarks are presented in Section 6. An appendix 
contains the details of the compactification lattice effect on the 
inflaton potential by summing over the images of a brane. 

\section{Effective Action for Brane-Stacks}

We write the world-volume action for $N$ parallel BPS D$p$-branes and $M$ 
anti-D$p$-branes with $d$ large extra dimensions.  This action
contains all the salient features of the more general effective action
which are important in the brane cosmology model: the ``slow-roll''
potential for the inflaton (brane separation) coming from the
combined R-R and NS-NS sectors, multiple and non-commuting scalar
fields, and a set of scalar fields which become tachyonic at the end
of inflation (or possibly after the slow roll phase has ended).  The
effective action is expanded about a flat background, with metric
$\eta = $diag$(-1,1,\ldots)$\footnote{The effects of the curved
background due to the branes is taken into account by a doubling of
the R-R sector force.}:
\baray\nonumber
S = -\tau_p\int d^{p+1}\sigma&&\left\{
\Tr{N}\left[\frac 1 2D^+_\mu\phi^I D^{+\mu}\phi^I
  + \frac 1 4(F^+)^2\right]
+\Tr{M}\left[\frac 1 2D^-_\mu \tilde\phi^I D^{-\mu} \tilde\phi^I
  + \frac 1 4(F^-)^2\right]\right.\\
&&\left.+\frac{1}{2} \Tr{N}\left[D^\pm_\mu T D^{\pm\mu} T^\dagger\right]
+V(\phi,\tilde\phi,T)\right\}.
\label{S2}
\earay
The brane tension is $\tau_p = \a^{-(p+1)/2}/g_s(2 \pi)^p$, the $N$ brane 
coordinates, $\phi^{I=1,\ldots,d}$, are scalars in the adjoint of $U(N)$ 
written as $N \times N$ matrices in the fundamental representation, 
$\tilde\phi^I$ are  $M \times M$ 
scalars in the adjoint of $U(M)$ representing the anti-brane coordinates.  
$T$ is in the $(\mathbf{N},\overline{\mathbf{M}})$ of $U(N)\times(M)$ 
\cite{k-theory}
and $F^{\pm}$ are the field strengths of $A^{\pm}$, the $U(N)$ and $U(M)$ 
connections on the brane and anti-brane stacks in the fundamental 
representations.  The gauge covariant derivatives are
\baray\nonumber
  D^+_\mu\phi^I &=& \partial_\mu\phi^I+\frac{i}{2\pi\a}[A^+_\mu,\phi^I],
\\\nonumber
  D^-_\mu\tilde\phi^I &=& \partial_\mu\tilde\phi^I+\frac{i}{2\pi\a}
[A^-_\mu,\tilde\phi^I],\\\nonumber
  D^\pm_\mu T &=& \partial_\mu T+\frac{i}{2\pi\a}(A^+_\mu T-TA^-_\mu),
\earay
The potentials are given by
\baray
\label{SVall}
V(\phi,\tilde\phi, T) &=& V_b(\phi,\tilde\phi) + V_s(\phi,\tilde\phi,T) + V_T+ 
    V_l(\phi,\tilde\phi),\nonumber\\ 
V_b &=& -\frac{1}{4(2 \pi\a)^2}\left\{\Tr{N}([\phi^I,\phi^J][\phi^I,\phi^J])+
    \Tr{M}([\tilde\phi^I, \tilde\phi^J][\tilde\phi^I, \tilde\phi^J])\right\},
\nonumber\\
V_s &=&  \frac{1}{2(2\pi\a)^2} \Tr{N}\left[\phi^ITT^\dagger\phi^I
+T\tilde\phi^I\tilde\phi^IT^\dagger
    -2\phi^IT\tilde\phi^IT^\dagger\right],\nonumber\\
V_T &=& -\frac{1}{4\a} \Tr{N}(TT^\dagger)+O(T^4),\nonumber\\\label{v3}
V_l &=& (N+M) - \sum\limits_{nk} \frac{\hat \beta g_s^2\a^4\tau_p}
{\left[\sum\limits_I(\phi^I_{nn} -\tilde\phi^I_{kk})^2\right]^{(d-2)/2}},
\earay
where $\hat \beta=2^5\pi^{7-d/2}\Gamma((d-2)/2)$.  The potential $V_b$
is obtained from the non-Abelian generalization of the Dirac
Born-Infeld action \cite{Myers}.  The form of the tachyon scalar
interactions, $V_s$, can be easily deduced when $N\ne M$, these being
the only quartic interactions allowed.  The coefficients can be
deduced by specializing to the $N=M=1$ case, and using the fact that
the tachyon quadratic term goes negative for $(\phi-\tilde\phi)^2 <
2\pi^2\a$ \cite{BanksSusskind}.  Alternatively these terms appear as
the T-duals of the tachyon gauge-kinetic terms.

The tachyon potential, $V_T$, can been calculated by various means 
\cite{Berk,BSZ,Kutasov}:
\begin{itemize}
\item  The result from level truncated cubic string field 
theory \cite{Berk,BSZ} gives a tachyon mass$^2$ which agrees with that of 
the NS open string tachyon, $-\frac1{2\a}$\cite{Polchinski1}; this is the 
mass of the perturbative state which appears in the brane-brane system upon 
reversal of the R-R charge of one.
\item The tachyon potential obtained from boundary string field 
theory (BSFT) \cite{Kutasov} is not in canonical form, and a field 
redefinition to its canonical form will involve all components of 
the string field.  BSFT does however have the advantage that it
clearly verifies Sen's conjecture \cite{Sen2} that at the minimum of the 
tachyon potential, the physics is that of the closed string vacuum.
\item The $T^4$ term in $V_T$ has been calculated at the lowest level in 
cubic string field theory for $N=M=1$ to be $+\frac1{2\a^2}T^4$ \cite{Berk}, 
but there is ambiguity in its form in general, with both $\Tr{}(T^4)$ and 
$[\Tr{}(T^2)]^2$ terms allowable; BSFT predicts only the former 
term \cite{Kutasov}, but it is 
uncertain whether this holds under the field redefinition.
\item For the purposes here, the $T^2$ term is most important, since its 
coefficient dictates the separation at which $T$ becomes tachyonic, ending 
inflation.\footnote{Were we to use the BSFT mass$^2$ of $-1/4\ln2\a$, the 
end of the inflationary phase would be changed little.}  In
generalizing this action to that of branes at angles, the coefficient
of $T^2$ term should become \cite{Polchinski1}
$-\frac{\theta}{4\pi\a}$, where $\theta$ is the angle between the
brane-stacks; $\theta=\pi$ reverts to the brane anti-brane case above.
The effects of placing the branes at angles is to lessen the distance
between the branes at which $T$ becomes tachyonic.
\item The presence of the tachyonic modes originates from the matrix 
nature of the brane positions. This non-commutative property has 
interesting cosmological consequences. The non-trivial vacuum structure 
of the tachyon condensate allows the creation of lower-dimensional branes. 
Since the tachyons are not the inflaton, lower-dimensional brane production
will take place after inflation, typically via the Kibble mechanism.
It is important that this does not re-introduce the old monopole problem.
\end{itemize}

Finally, the potential $V_l$ takes into account the R-R and NS-NS 
backgrounds generated by the brane and anti-brane stacks
\cite{Polchinski2,BanksSusskind}.  $\phi^I_{ln}$ is the $ln$ component
of $\phi^I$, and $V_l$ is a function of $\phi^I_{nn}
-\tilde\phi^I_{kk}$ only.  For $d=2$, the Coulomb-like potential is
replaced by a log.  This is twice the Chern-Simons brane action,
written with the explicit expression for the pull-back of the
space-time $p+1$ form potential \cite{Polchinski2} generated by the
branes.  Only the term from the massless modes in the expansion of the
exact potential is necessary for our purposes, to give the
long-distance potential between the brane stacks, before the $T$
fields become tachyonic, and inflation ends.  We have doubled the R-R
potential to account for the equal force supplied by the NS-NS
sector.  When we consider branes at angles, the coefficient of $V_l$
acquires a factor of $\tan(\theta/2)\sin^2(\theta/2)$
\cite{Polchinski1,rabadan1}, with more complicated angular factors for
more than one tilted direction; as we shall see,
some brane inflation scenarios require
branes at small angles to achieve sufficient e-foldings.

\section{Braneworld at the Start of Inflation}

\subsection{Set-Up}

Realistic string models have 6 of the 9 space dimensions compactified.
Consider $Dp$-branes in 10-dimensional space-time, where $(p-3)$
dimensions parallel to the brane are compactified with volume $\Vpl$ and 
the $d$ dimensions orthogonal to the brane are compactified with volume 
$V_d$. The remaining $3$ spacial dimensions of the $Dp$-brane are 
uncompactified; they span our observable universe.
Let $M_s$ be the string scale $M_s^{-2} = \alpha^{\prime}$.
So $4 +(p-3) + d=10$ and $V=V_d\Vpl$. In general, the branes in 
the early universe that will eventually collide after inflation but
before the big bang are different from those present in 
today's braneworld, so $\Vpl$ can be quite different from $V_t$ 
in Eq.(\ref{MP1}). To simplify the discussion, we shall assume, unless 
pointed out otherwise, that they largely overlap. 

In string theory, there is a T-duality symmetry, {\em i.e.}, physics 
is invariant under a T-duality transformation.
For toroidal compactification, we have 
$V = \prod_i l_i = \prod _i 2 \pi r_i$, where $l_i$ is 
the size of the $i$th torus/circle and $r_i$ the corresponding radius,
$l_i= 2\pi r_i$.\footnote{In general, the radius $r_i$ here does not 
necessarily have to be the radius of the $i$th torus. It is
simply a characteristic length scale of the compactified dimension.
In the case of a ${\bf Z}_N$ orbifold, the volume is given
by $\prod_i v_i={1\over N} \prod_{i} (2 \pi R_i)^2  \equiv \prod_{i}
(2 \pi r_i)^2$. In general, it stands for the first KK mode, 
{\em i.e.}, at $r_i^{-1}$.} 
If any of the $r_i$ is much smaller than the string scale, 
{\em i.e.}, $M_s r_i <1$, the T-dual description is more appropriate :
\be
         g_s \rightarrow 
         {g_s \over {r_i M_s}}, \quad \quad 
         r_i \rightarrow   {1 \over {r_i M_s^2}}
\ee
In this dual picture, the new $r_i M_s$
of the dual $T_i$ torus is always larger than or equal to unity.
Under this duality transformation, the Dirichlet and Neumann
boundary conditions of the open strings are interchanged, and so
the branes are also mapped to other types of branes. 
This allows us to consider only the cases where $M_sr_i \ge 1$, with
$M_sr_i = 1$ the self-dual point.

\begin{itemize}
\item Perturbative string theory does not seem to stabilize the dilaton.
So we expect dilaton stabilization to arise from non-perturbative 
stringy effects. Very strongly coupled string theory presumably has 
a weak dual description, where the dilaton is again not stabilized.
This leads us to conclude that the string coupling generically is 
expected to be 
$g_s {\ \lower-1.2pt\vbox{\hbox{\rlap{$>$}\lower5pt\vbox{\hbox{$\sim$}}}}\ }1$.
To obtain a theory with a weakly coupled sector in the low energy effective 
field theory (i.e., $\alpha(GUT)$ is small), it then seems necessary 
to have the brane world picture \cite{zura}.
Let us consider the $p>3$ case. In this case, 
\be
g_s= 2 (M_s\rpl)^{p-3} \alpha(\rpl) = 2 v_{\parallel} \alpha(\rpl) 
\ee
where $\vpl= (M_s \rpl)^{p-3}$ and $\alpha(\rpl)$ is the gauge coupling at 
the scale $1/\rpl$. All couplings should unify at the string scale.
To get a qualitative picture of the impact of a relatively large 
string coupling, let us take $g_s \sim 1$.
So $\alpha$ has logarithmic running up to the scale $1/\rpl$ and then
power-running between $1/\rpl$ and $M_s$, yielding an
$\alpha(M_s) \sim 1$.
For $p=5$, $M_s\rpl \sim 5$ while for $p=7$, $M_s\rpl \sim \sqrt 5$. So
$\alpha(\rpl)$ is essentially $\alpha_{GUT}$. Since the renormalization 
group flow $\beta$ coefficients (of the standard model gauge couplings)
for power-running \cite{dienes} have almost identical convergence 
properties as that for the logarithmic running in MSSM and the 
power running is only over a relatively small energy range, the
unification of the gauge couplings at the string scale is probably
assured.
\item Of the $d$ dimensions orthogonal to the brane, suppose 
only $\dper$ of them are large. 
To simplify the discussions, let the 6-D compactification volume $V$ be
\be
V=\Vpl \Vp V_{(d-\dper)}  
\ee
where the $(d- \dper)$-D volume $V_{(d-\dper)}$ is $(2\pi/M_s)^{d-\dper}$
(i.e., at the self-dual value). We shall consider $\dper + p \le 9$. 
Unless pointed out otherwise, we shall 
assume all the large extra dimensions to have equal size $l_i=\lp$ or
$r_i=\rp$ ($\lp=2 \pi \rp$), so the remaining dimensions orthogonal to 
the brane have sizes such that $M_sr_i=1$. For branes at angle and
$p>3$, we shall consider $\dper$ to range from 2 to 4.
  
Note the effect of a larger $V_{d-\dper}$ is equivalent to an increased 
$\dper$, so the physics will simply interpolate between the values 
considered. 
This allows us to use the volumes
$\Vp = (2 \pi\rp)^{\dper}$ (transverse to the brane) and $\vpl$
(along the brane) as parameters to be determined.
Following from dimensional reduction where the 10-D 
Newton's constant $G_{10}$ is given by Eq.(\ref{g10}), we have
\be
\label{MPdef}
g_s^2 M_P^2 = M_s^2 (M_s\rp)^{\dper} \vpl /\pi
\ee

Note that the factor of $(2 \pi)^6 \pi$ in Eq(\ref{MPdef}) is 
missing in Ref.\cite{dvali,burgess,rabadan}.
The string coupling is $g_s = e^{\phi_D}$ 
where $\phi_D$ is the dilaton.
The BPS D$p$-brane has brane tension $\tau_p$ and R-R charge $\mu_p$:
\be
\tau_p= \frac{M_s^{p+1}}{(2\pi)^p g_s}, \quad \quad \mu_p = g_s\tau_p
\ee
while an anti-D$p$-brane has the same tension $\tau_p$ but opposite 
R-R charge. These relations are subject to quantum corrections, which 
we shall ignore for the moment. 

We shall demand that $M_s \rp \gg 1$ and $\dper > 1$.
For instance, this is required within string theory if we want to
treat the bulk dynamics using only the low-energy effective field 
theory corresponding to the massless string states. These
are also the conditions under which the nonlinear contributions of
Einstein's equations are negligible when considering the gravitational
field of a single $D$-brane, say. Therefore our approximate
4D effective field theory treatment is self-consistent. 
\item In principle, we can also consider $\dper=1$. However, in this 
case, the potential between branes becomes confining and the presence 
of branes will induce a warped geometry in the bulk \cite{RS}. 
As a result of this, the physics is quite different and will not be 
considered here. 
\item The standard model of strong and electroweak interactions 
requires at least 5 $Dp$-branes. Suppose today's universe is described 
by a Type I or an orientifold string model, which is supersymmetric at 
scales above the electroweak scale (or non-super\-symmetric but with a 
very small cosmological constant).
Suppose the universe starts out with more branes than its vacuum state.
This may happen if the early universe is very hot, which probably is
a hot gas of different $p$-branes oriented randomly.
Alternatively, the universe may start out with some
brane-anti-brane pair or non-BPS branes \cite{Sen1,Sen2}. 
This set-up may emerge without a very hot universe.
Eventually, the branes will move and collide, until it reaches the
brane configuration of the vacuum state; that is, 
all branes except those in the ground state of the string 
model disappear, via collisions, decays and/or annihilations. 
Before the last collision, we want to see if inflation happens, 
and how generic inflation is. If the branes that will collide 
are randomly placed in the compactification volume, we want to
estimate the probability {\it P} that sufficient inflation will take 
place before collision. The universe will then reheat to 
originate the big bang. When this happens, the magnitude of the 
density perturbation will be used to fix the string scale.
It is important to see if the density perturbation power 
spectrum index satisfies the observational bound.
\end{itemize}

\subsection{Brane-Anti-Brane Pair}

Suppose that inflation takes place before the last  
$Dp$-brane-anti-$Dp$-brane pair annihilates.
We may also consider the case of a stack of parallel $N$ $Dp$-branes 
and a stack of $M$ parallel anti-$Dp$-branes. The 4-D inflaton potential has a 
term due to brane tension and a term due to the interactions 
between the branes. The force between stationary parallel BPS $Dp$-branes 
is zero, due to the exact cancellation between the attractive NS-NS and 
the repulsive Ramond-Ramond (R-R) couplings. On the other hand, the 
potential between a $Dp$-brane and an anti-$Dp$-brane is attractive, since 
the R-R coupling is also attractive in this case. 
To simplify the problem, let us take $N=M=1$, i.e., a single anti-$Dp$-brane.
Inflation takes place when the last anti-$Dp$-brane 
is approaching a $Dp$-brane and inflation ends when it annihilates with the
brane. The universe is taken to be supersymmetric at this scale, so 
after annihilation, we expect the vacuum energy 
to be essentially zero, where the remaining brane tensions are exactly 
canceled by the presence of orientifold planes. We shall consider the 
simple case where the remaining branes are sitting at the orientifold 
fixed points (evenly distributed). In this case, the interaction of the 
anti-$Dp$-brane with the remaining branes is exactly canceled by that with 
the orientifold planes. 
The 4-D effective action of a pair of $Dp$-anti-$Dp$-branes
takes the form:
\be
\label{effective}
\Gamma  \simeq \int d^4 x \sqrt{|g|} 
\left( \frac{\tau_p\Vpl}{2} (\partial_{\mu} y_b \partial^{\mu} y_b
+\partial_{\mu} y_a \partial^{\mu} y_a)
-V(y) + \dots  \right) \nonumber \\
\ee
where the brane is at $y_b$ and the anti-brane is at $y_a$, so the 
separation is $y= y_b-y_a$. With effective $\kappa^2/(\Vpl V_{d-\dper})$ 
where $V_{d-\dper}= (2\pi/M_s)^{9-p-\dper}$, 
the potential for $\dper>2$ is easy to write down :
\baray
\label{poten}
V_l(y) = 2 \tau_p\Vpl - \frac{\kappa^2 \beta \tau^2_p \Vpl }
{y^{\dper-2}} \left(\frac{M_s}{2 \pi}\right)^{9-p-\dper}
\earay    
where
\baray
\tau_p &=& M_s^{p+1}/(2 \pi)^p g_s, \nonumber \\ 
\beta &=& \pi^{-\dper/2}\Gamma((\dper-2)/2)/2 \quad \dper>2
\earay
and $\kappa^2$ is given in Eq.(\ref{g10}). 
Here $\beta/2$ comes from the inverse surface area from Gauss's Law and the
extra factor of $2$ in $\beta$ is due to the sum of the NS-NS 
and the R-R couplings. The potential becomes logarithmic 
and $\beta=1/\pi$ for $\dper=2$. 

\subsection{Branes at Angles}

String models with branes at angles can be quite realistic \cite{angle}.
Following Ref.\cite{rabadan}, let us consider a supersymmetric string 
model (or a non-supersymmetric model with a zero cosmological constant) 
where a brane wrapping 1-cycles in a two dimensional torus. Let the
two torus have radii $\rpl$ (or size $\lpl = 2 \pi \rpl$) and $u\rpl$, 
where $u <1$. 
This brane wraps 
a straight line in the $\{n_f [a] + m_f [b]\}$ homology class. 
The energy density of this brane is just \cite{Polchinski1}
\be
E_f = \tau_4 \lpl \Big[\sqrt{n_f^2 + (u m_f)^2} \Big]
\ee
This energy is of course precisely canceled by the presence of other 
branes and orientifold planes in the model.
Suppose this brane results from the collision of two branes in the 
early universe, each of which wraps a straight line in the
$\{n_i [a] + m_i [b]\}$ homology class. Since the total (homological) 
charge is conserved, $(m_f, n_f)= (m_1 + m_2, n_1 +n_2)$. 
The energy density of the two-brane system is 
$$E_2 = \tau_4 \lpl \Big[\sqrt{n_1^2 + (u m_1)^2} + \sqrt{n_2^2 +
(u m_2)^2}\Big]\,.$$ 
So the energy density before the brane collision, up to an interaction 
term, is $$V_0=E_2 -E_f$$ Let us consider the case where $V_0>0$.
Before collision, the branes are at an angle  $\theta=\phi_1-\phi_2$,
where $\tan \phi_i =  (u m_i/n_i)$ and are separated by a distance in 
directions orthogonal to the torus.
When this separation between the branes 
approaches zero (see Figure 3), the collision results in the
$(m_f, n_f)$ brane.
In the case where $(m_f, n_f)= (0,0)$, with non-zero $(m_i, n_i)$,
we have a brane-anti-brane pair (with $\theta=\pi$) which annihilate 
when they collide. When $(m_f, n_f)$ is non-zero, a brane is left 
behind. If $m_f$ and $n_f$ are co-prime, this brane is at an angle. 
Such branes appear naturally in many phenomenologcally interesting 
models \cite{angle}. If $m_f$ and $n_f$ have a common factor $k$, 
$m_f=k\hat m_f$, $n_f=k\hat n_f$, we may consider this as $k$ branes 
in the $\{\hat n_f [a] + \hat m_f [b]\}$ 
homology class. If either $m_f$ or $n_f$ is zero, presumably the 
brane (or branes) is parallel to an orientifold plane.
In effective low-energy field theory, this is the Higgs mechanism.
Suppose there is a $U(N)$ gauge group associated with each brane
before collision (i.e., $n_1=n_2=N$).
Then there is a $U(N) \times U(N)$ gauge symmetry together with 
a bifundamental scalar, which are open strings stretched between 
the branes. During collision, the bifundamental scalar develops a
vacuum expectation value and spontaneous symmetry breaking takes place:
\be 
\label{higgs}
U(N) \times U(N) \rightarrow U(N)
\ee 

\EPSFIGURE[t]{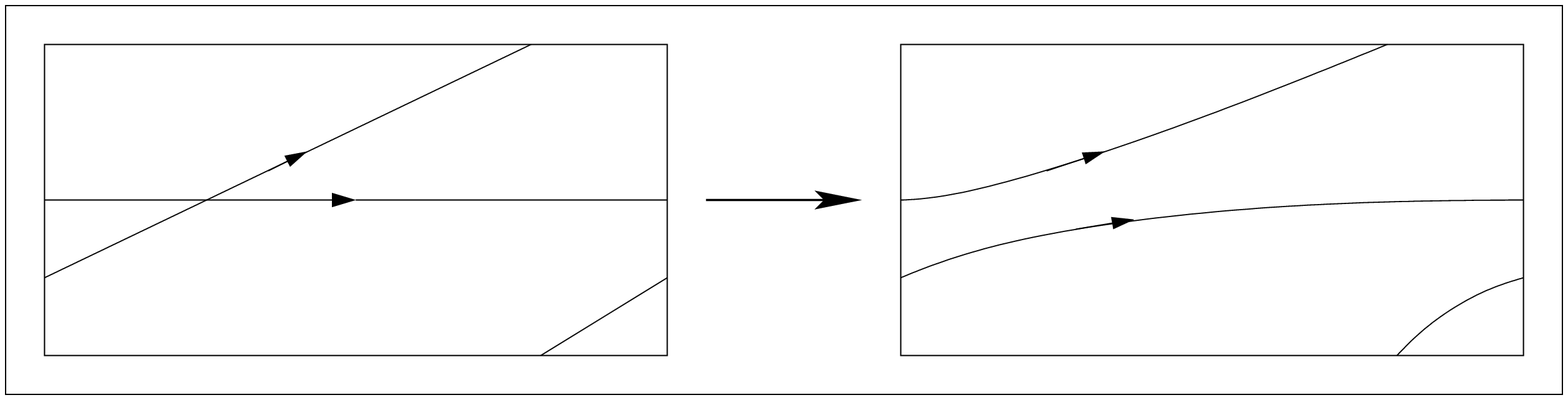,width=1.0\textwidth} 
 {Branes wrapping a torus with sides $\lpl$ and $u \lpl$, where $u<1$.
The two branes ((1,1) and (1,0)) are at an angle $\theta$, but they are 
separated in directions orthogonal to the torus. When that separation 
approaches zero, the two branes collide to become a single (2,1) brane. }

Suppose these two branes are separated by a large distance in the other 
compactified directions. 
Then the effective potential is given by \cite{Polchinski1,rabadan1}
\be
\label{Vata}
V(y) = V_0 - \frac{\beta M_s^{6-\dper}\sin^2\theta/2\,\tan\theta/2 }
{4\pi y^{\dper -2}}.
\ee
where $V_0 \propto \tan^2 \theta$. 
For $\dper=2$, the potential has a logarithmic form.
Note that the potential term vanishes 
as $\theta  \rightarrow 0$. In this limit, the branes are parallel and 
are BPS with respect to each other. In the other limit, as 
$\theta  \rightarrow \pi$, the potential blows up due 
to the $\tan\theta/2$ factor. This happens because the brane becomes 
parallel to the anti-brane and so the strings stretching between them are 
free to move in that direction. In this case, there is a volume factor
to be taken into account. Alternatively, one simply use the  
potential for the brane-anti-brane system given earlier. 

Here, we are interested in the small angle case. To get a qualitative 
feeling of the scenario, let us take for example  
$m_1=1$ and $m_2=0$. Then
$\phi_2={\rm arctan}(um_2/n_2)=0$, and  
$\phi_1={\rm arctan} (u/n_1)\equiv \theta \ll 1$. 
This may be achieved with large $n_1$ or small $u$, or a combination 
of both. 
Small $\theta$ may be realized in a number of ways in string theory :

$\bullet$ We may simply take $n_1$ to be very large, the case considered 
in Ref.\cite{rabadan}.
Generically, large $n_1$ implies the string model today will have 
a brane with a very high winding number. This is a possibility that 
deserves further study. 

$\bullet$ Suppose the resulting brane that wraps around $\lpl$ is not 
among the branes that are responsible for the standard model gauge 
fields, that is, $\lpl$ is part of $\Vp$ and not part of $\Vpl$. 
Then there is hardly any constraint on the value $\lpl$ and we may 
take $u$ to be very small with $n_1=1$, say.   

$\bullet$ Generically, the resulting brane will wrap around $\lpl$, so
$\lpl$ is likely to be a part of $V_t$ in Eq.(\ref{MP1}).
We may estimate how small $u$ may be. 
First recall that $uM_s\lpl \ge 2 \pi$. On the other hand, $\lpl$ is 
bounded by $M_s\lpl/{2 \pi} \lesssim \vpl$, so, 
following Eq.(\ref{gsalpha}), $u \ge 2 \alpha/g_s$.
The smallest we can reasonably have is $u \approx 10^{-2}$.
Since it is quite reasonable for a string model to have $n_1 \ge 1$
(to accomodate the standard model, we expect $n_f \ge 5$, which implies 
$n_1 \ge 4$ in our example),
$\theta \simeq u/n_1 \approx 10^{-2}$ is not unreasonable.
As we shall see, $\theta \sim 0.1$ is quite sufficient.

To facilitate comparison with Ref.\cite{rabadan},
we may consider the $\dper =4$ case.
In the small angle approximation we have (taking $n_1=n_2=1$),
\be
\label{Vatsa}
V(y) = \frac{\tau_4\lpl \tan^2\theta}{4}  - 
\,{M_s^2\sin^2\theta/2\,\tan\theta/2 \over8\pi^3 y^2}\,.
\ee

\section{Inflation}

\subsection{General Set-Up}

The inflaton as seen by a 4-D observer 
\be
\Gamma \simeq \int d^4 x \sqrt{|g|} 
\left( \frac{1}{2} \partial_{\mu} \psi \partial^{\mu} \psi
-V (\psi) +...  \right) 
\ee
is related to the brane separation $y=y_b-y_a$ in Eq.(\ref{effective}) 
by:
\be
\psi = y\sqrt{\frac{\tau_pV_{\parallel}}2}=yM_s^2\sqrt{\frac{\vpl}
{2g_s\left(2\pi\right)^3}}
\ee
where $\vpl=\left(M_s^{p-3}V_{\parallel}\right)/\left(2\pi\right)^{p-3}$.
So the effective potential for the inflaton becomes, 
for $\dper>2$ and $\dper=2$, respectively,
\baray 
V(\psi) &=&  A \left( 1 - \frac{\lambda}{\psi^{\dper -2}}\right)
\nonumber \\ 
V(\psi) &=&  A \left( 1 + {\lambda} \ln (\psi/M_s) \right)
\earay
where $A$ and $\lambda$ are given by Eq.(\ref{poten}) for the 
brane-anti-brane case and by Eq.(\ref{Vata}) for the branes at angle case. 
For small $\theta$ and $\dper=4$, we shall use Eq.(\ref{Vatsa}).

The equation of motion for $\psi$ is
\begin{equation}\label{eom}
 \ddot{\psi} + 3 H \dot{\psi}+ V^{\prime} = 0
\end{equation}
where the prime indicates derivative with respect to $\psi$.    
For inflation to take place, there is some choice of $\psi$ so that the
potential $V(\psi)$ satisfies the slow-roll conditions $\epsilon << 1$ and 
$|\eta | << 1$, where :
\baray
\label{slow-roll}
\epsilon &=& \frac{1}{2} M_P^2 \left( \frac{V^{\prime}}{V} \right)^2 
 \nonumber \\
\eta  &=&  M_P^2 \frac{V^{\prime\prime}}{V}
\earay
During the slow-roll epoch, $\ddot{\psi}$ in Eq.(\ref{eom}) is negligible 
and the Hubble constant $H$ is given by 
\be
H^2 \simeq \left(\frac{\dot{a}}{a} \right)^2 = \frac{V}{3 M_P^2}
\ee
where $a(t)$ is the cosmic scale factor, and
we may relate the value of $\psi_N$ as a function of the number 
$N_e$ of e-foldings before the end of inflation.
The value of $\psi_{end}$ is determined when the slow-roll condition
breaks down, when either $ |\eta| \approx 1$ or $\epsilon \approx 1$. 
Generically, $ |\eta| >> \epsilon$ in brane inflationary scenarios. 
It takes $N_e$ e-foldings for $\psi_N$ to reach $\psi_{end}$.
\be
N_e= \int_{t_N}^{t_{end}} H dt=
\int_{\psi_{N}}^{\psi_{end}} H \frac{d \psi}{\dot{\psi}}
=\frac{1}{M_P^2} \int_{\psi_{end}}^{\psi_N} \frac{ V }{V^{\prime}}
 d \psi
\ee
Inflation may also end 
when the ground state open string mode becomes tachyonic: 
\be
m^2_T= \left(\frac{M^2_sy}{2 \pi} \right)^2 - \frac{M_s^2\theta}{2\pi}
\ee
However, in the scenarios we are considering, the slow-roll condition 
breaks down before the tachyon mode appears. As we shall see, this has 
a non-trivial impact on defect productions after inflation.

We know that the density perturbation measured \cite{cobe} from Cosmic 
Microwave Background (CMB) is $\delta_H = 1.9 \times 10^{-5}$.
This is the density perturbation at $N_e$ e-foldings before the end 
of inflation, where 
\be
\label{Ne}
N_e \simeq 60 + \frac{2}{3}\ln(M_s/10^{16}GeV)+
\frac{1}{3}\ln(T_{RH}/10^{14}GeV)
\ee 
where $T_{RH}$ is the reheating temperature after inflation.
Here, for $\dper \ge 2$,
\be
 N_e = \frac{\dper-1}{\dper}\left[\frac{y_N}{y_{end}}\right]^{\dper}
\ee
The power index of the density perturbation 
\be
n-1= 2 \eta -6 \epsilon
\ee
For $\epsilon \ll \eta$, the spectral index $n$ and its variation with 
respect to the wavenumber $k$ are given by:
\be
n-1 = \left.2\eta\right|_{N_e} \approx -\frac{2\left(\dper-1\right)}
{\dper N_e} 
\ee
\be
\frac{dn}{d\ln k} = \left.2\xi^2\right|_{N_e} = 2M_p^4\frac{V^{\prime}
V^{\prime\prime\prime}}{V^2} \approx
\frac{2\left(\dper-1\right)}{\dper N_e^2}
\ee
In our case, we have
\baray
\eta &=& - \beta (\dper-1)(\dper -2)\left(\frac{\lp}{y} \right)^{\dper}  
 \quad  \quad  \dper>2  \nonumber \\
&=& -\frac{1}{\pi}\left(\frac{\lp}{y} \right)^2 \quad \quad \quad \dper=2
\earay
Generically, we need $|\eta|  \lesssim 1/{60}$ during the early inflationary 
stage. Since $ |y_i| < \lp/2$, we see that $\eta$ can never be small enough 
to satisfy the slow-roll condition. This is noted in 
Ref.\cite{dvali,burgess}.

Now let us consider the more general situation of branes at angle.
As pointed out in Ref.\cite{rabadan}, the situation improves.
For small $\theta$ and $n_1 = 1$, we have $\theta \simeq u/n_1=u$.
With $(2 \pi)^{\dper +2} \pi g_s^2M_P^2 = 
M_s^2 (M_s\lpl)(M_s u\lpl)(M_s \lp)^{\dper}$,
\baray
\label{angeta}
\eta &\approx& -\frac{3 \theta^2}{2 \pi^2} \left(\frac{\lp}{y}\right)^4
 \quad \quad  \dper=4 \nonumber \\
&\approx&  -4 \pi \theta^2  \left(\frac{\lp}{y}\right)^2
 \quad \quad  \dper=2 
\earay
To have enough number of e-foldings of slow-roll while the branes 
are separated, we get a bound on $\theta$ by requiring $|\eta| \le 1/{60}$
at the beginning of inflation.
However, when $|y| \sim \lp/2$, the compactification effect is important 
and Eq.(\ref{angeta}) for $\eta$ is not valid (see the discussion below). 
For the compactification effect to be negligible so Eq.(\ref{angeta}) is 
valid, a smaller $\theta$ is needed, say  
\be
\theta  < 10^{-2}
\ee
With such a small $\theta$, 
the probability ${\it P}$ of sufficient inflation approaches unity.
A small but not too small value of $\theta$ will still increase 
${\it P}$ and further improves the naturalness 
of existence of an inflationary epoch in the brane world. 

\EPSFIGURE[b]{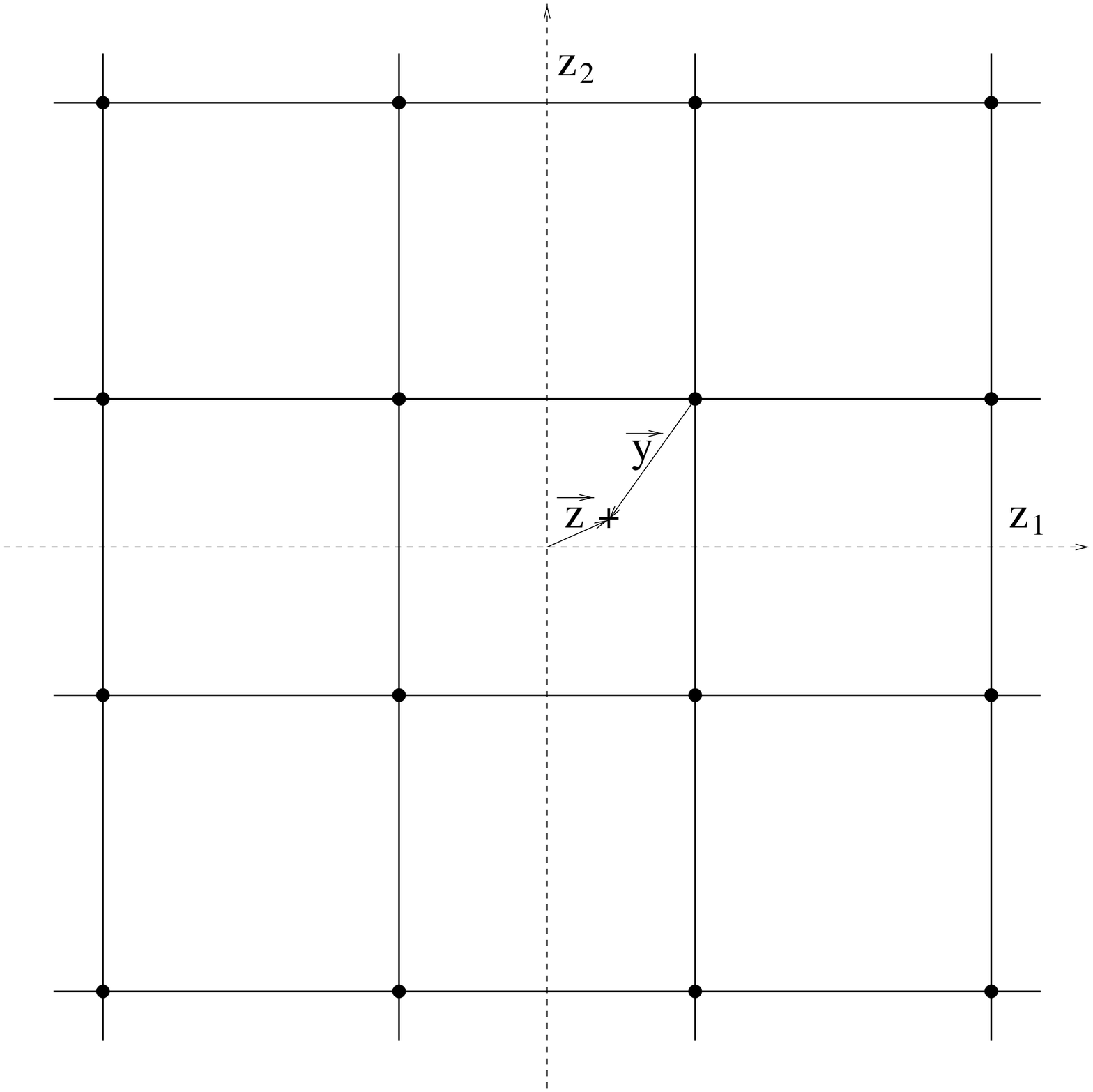,width=0.4\textwidth}
 {Compactification on hypercubic lattice (solid lines), where one 
brane sits at the lattice points while the other brane is at $z$. 
Here, $z$ (and $\varphi$) are measured with respect to the antipodal 
point (the intersection of the dashed lines).}

\subsection{Hypercubic Compactification}

When the brane separation
is comparable to the lattice size, we must include the forces
exerted by the images of one brane on the other brane.
(The net force on a brane due to its own images is exactly zero.) 
This always tend to soften (or flattens) the potential. 
Let us first consider the hypercubic lattice for the brane-anti-brane 
pair discussed in Ref.\cite{burgess}. Let the brane sit at the 
origin $\overrightarrow y = \overrightarrow 0$,
while we measure the position of the anti-brane from the antipodal point 
$(\lp/2, \lp/2, ...)$, at which the net force on the anti-brane is zero 
: $z_i = y_i - \lp/2$ and define the corresponding 
$\varphi_i=\psi_i - \psi_i*$, where $\psi_i*$ is the value of 
$\psi_i$ at the antipodal point. (See Figure 4.)
The potential for the D-dimensional hypercubic lattice is given by:
\be
V\left(\overrightarrow y \right) = \left\{ \begin{array}{l}
A - B\sum_{i}\frac{1}{\left|\overrightarrow y -\overrightarrow r_i 
\right|^{\dper - 2}}, \quad \dper > 2 \\
A + B \sum_{i}\ln\left|\overrightarrow y -\overrightarrow r_i \right|, 
 \quad \dper = 2
\end{array}
\right.
\ee
where the sum is over all the lattice sites $\overrightarrow r_i$
(positions of the images of the brane), 
where $\overrightarrow r_0$ is at the origin 
$\overrightarrow y = \overrightarrow 0$.
In order to estimate the potential around the antipodal point,
we expand to 4th order in $\overrightarrow z$ (or $\varphi$) around 
the center of 
the elementary cell, i.e., $\overrightarrow z = \overrightarrow 0$.
Summing over the images of the other brane for hypercubic lattices
(see Appendix for details), we have, for small $\varphi$, 
\be
\label{Vhc}
V(\varphi) \simeq A \left( 1 - \hat \lambda  [ D\sum_{i \ne j} 
\varphi_i^2\varphi_j^2 - C\sum_i \varphi_i^4] \right)
\ee
where $\hat \lambda$ is related to $\lambda$ by some rescaling,
$$\hat \lambda = \lambda \lp^{-2-\dper} 
\left( \frac{\tau_p \Vpl}{2} \right)^{-1 -\dper/2} $$
and the values of $C$ and $D$ are given in Table 1.
As pointed out in Ref.\cite{burgess}, the quadratic terms in $\varphi$ 
are absent in hypercubic lattices. 

\begin{center}
\begin{tabular}{|c|r|r|r|r|r|r|}
\hline
$\dper$ & C~~ & D~~ & F ~~ & $F\beta$~~& $2z_{*}/\lp$~ & ${\it P}(\pi)$~ \\
\hline\hline

2 & 3.94 & 23.64 & 15.76 & 5.0 & 3.5 \% & 2$\times 10^{-3}$ \\
\hline 

3 & 6.22  & 18.64  & 37.26 & 5.9  & 3.3 \%  &  $<10^{-4}$ \\
\hline

4 & 15.41 & 30.82  & 123.3  & 6.2   & 3.1 \% &  $\sim 10^{-6}$ \\
\hline 

\end{tabular}
\end{center}

The inflaton is a multi-component field, so the inflaton path can be 
quite complicated. Some sample paths are shown in the figures.
To get a qualitative feeling of the potential, let us consider the 
diagonal path
\baray
V(\varphi) &\simeq& A \left( 1 - \hat \lambda F \varphi^4 \right) \nonumber \\
F &=&  \frac{\dper(\dper -1)}{2}D  - \dper C
\earay

\DOUBLEFIGURE[t]{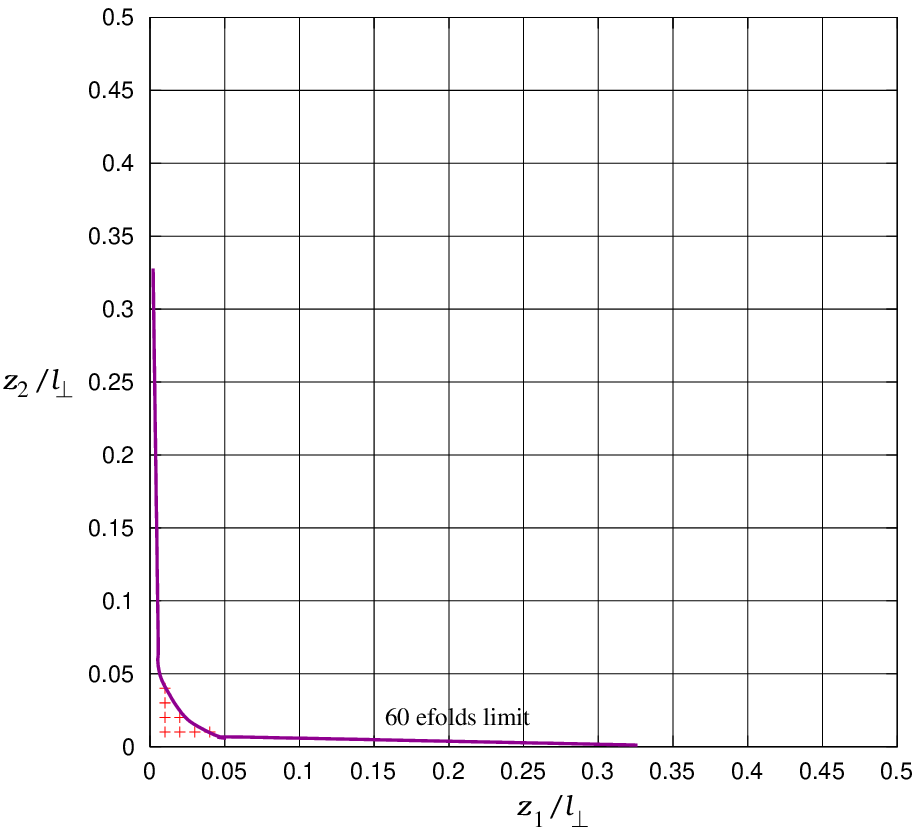,width=0.4\textwidth}
		{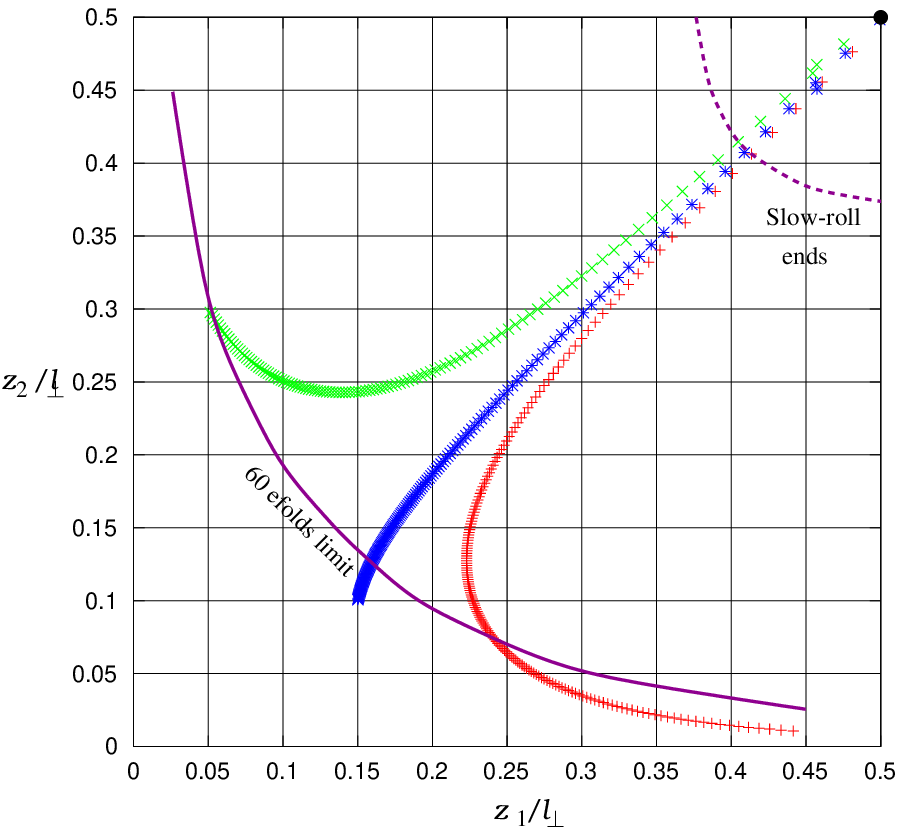,width=0.4\textwidth}
		{The 60 e-folds region for brane anti-brane interaction.
The brane is sitting at the upper right corner. There will be 
60 e-folds or more if the anti-brane starts in the region in the lower 
left corner.}
		{The 60 e-folds region for branes at fixed angle 
$\theta \sim 1/6$. Three possible inflaton paths with enough e-folds
are shown. Intervals along each path acts as a clock. Note that the 
right path starts out with $|\eta | \sim 1$ and then slows down during 
the inflationary epoch.} 
For large $N_e$, we have 
\baray
\eta &=& -12 M_P^2 \hat \lambda F \varphi^2 = -12 F \beta
\left( \frac{z}{\lp} \right)^2  \nonumber \\
N_e &\approx& \frac{3}{2|\eta|}
\earay  
The quantum fluctuation in deSitter space $H/{2 \pi}$ yields
\be
\label{deltac}
\delta_H^2 =\frac1{75\pi^2M_P^6}\left.\frac{V^3}{\left(V^{\prime}\right)^2}
\right|_{\varphi_N} \approx \frac{32 A \hat \lambda F N_e^3}{75 \pi^2}
\ee
To have enough time for slow-roll, let us consider the value $z*$ at 
$|\eta | \approx 1/{60}$, which 
implies $$\frac{2z_{*}}{\lp} \approx 3\%$$ This means the 
probability ${\it P(\pi)}$, which is essentially the fraction of the 
cell volume that yields 60 or more e-foldings for $\dper \ge 2$,
\be
{\it P(\pi)} \sim \left(\frac{2z_{*}}{\lp} \right)^{\dper} 
\sim (3\%)^{\dper} 
\ee
of enough inflation for an initial randomly placed brane-anti-brane 
pair is quite small. A more careful estimate of the region with 
enough inflation in the $\dper=2$ case is shown in Figure 5 
(since the lattice has at least a 4-fold symmetry, we need to show only 
a quarter of a lattice cell in the $\dper=2$ case),
while the value of ${\it P(\pi)}$ is given in Table 1. In this case, 
a fine-tuning on the initial condition seems to be required to obtain 
enough inflation.

Fortunately, the situation is immensely improved when the branes are 
at a small $\theta$.
Actually, one may argue that, to have a brane-anti-brane pair 
(that is $\theta = \pi$) involves some sort of a fine-tuning because, 
apriori, randomly placed branes will be at an angle $\theta \ne \pi$.
To get a qualitative feeling of the impact of small $\theta$,
let us consider a 4-brane pair with $n_1=n_2=1$ in $\dper=$2, 3 and 4.
In this scenario, $\varphi^2 \approx \tau_4 \lpl z^2 /2$, and
\baray
\label{anglat}
\eta \simeq  - 6 \theta^2 F \beta \left( \frac{z}{\lp} \right)^2  
\earay
This is useful since $ F \beta$ is relatively insensitive 
to $\dper$.
The probability ${\it P(\theta)}$ for branes at small angle $\theta$ 
is related to that for the brane-anti-brane case via 
\be
 {\it P}(\theta) \sim  \left(\frac{\sqrt{2}}{\theta}\right)^{\dper} 
{\it P}(\pi) \sim \left(\frac{4.5 \%}{\theta} \right)^{\dper}
\ee
so ${\it P(\theta)}$ is of order unity for $ \theta \lesssim  1/{10}$.
In Figure 6, we show the case where $\theta \sim 1/6$ and $\dper=2$.
For the diagonal path, the dependence of $n$ of $N_e$ is essentially 
the same as in other scenarios where $\eta \gg \epsilon$,
\be
 n \simeq 1 - \frac{3}{N_e}  \quad \quad  \frac{dn}{d {\ln k}} 
\simeq - \frac{3}{N^2_e}
\ee
Notice that all inflaton paths (see Figure 6) tend to move towards 
the diagonal path. So we expect only at most a slight correction in 
$n$ and $\delta_H$ for non-diagonal paths. 

\subsection{The String Scale}

To see the relation between the string scale $M_s$ and $\delta_H$, let us
consider a particularly interesting scenario.
A phenomenologically interesting orientifold construction, either 
supersymmetric or non-supersymmetric, typically starts with 3 tori.
For small $\theta$, we already have 3 sizes : $\lpl$ and $\theta \lpl$ and  
$\lp$. This fixes the 3 torus sizes. Since the brane resulting from 
brane collision must wind around $\lpl$, we may choose, in 
Eq.(\ref{gutvalue},\ref{gsalpha}),  $\lpl=V_t$ for 4-branes, or 
$V_t = \theta \lpl^2$ for 5-branes, where 
$(2 \pi)^6 \pi g_s^2M_P^2 = M_s^2 (M_s\lpl)^2(M_s \theta \lpl)^2
(M_s \lp)^2$. Assuming that $\lp \gg \lpl$, we have effective $\dper=2$.
So using Eq.(\ref{gsalpha}), Eq.(\ref{deltac}) for the 
5-brane case becomes 
\be
\delta_H^2 \simeq  \frac{2 g_s \theta^4 F \beta  N_e^3}{ 75 \pi^3 
 \alpha^3} \left( \frac{M_s}{M_P} \right)^4
\ee
Using $\delta_H \sim 1.9 \times 10^{-5}$, $F \beta \simeq 5$ from Table 1, 
$N_e \simeq 60$, $g_s \sim 1$, $\alpha \simeq 1/{25}$ and 
$\theta \sim 2 \alpha$, we have
\be
M_s \simeq 2 \times 10^{15} GeV
\ee                 
Here, if $\theta$ is not small, then the probablity to have 
enough inflation, $\it P$, will be too small.
On the other hand, $\theta$ cannot be much smaller than $2 \alpha$, 
otherwise $V_t$ (and so $g_s$ in Eq.(\ref{gsalpha})) will be too big. 
It is precisely this range of $\theta$ that allows the compactification 
lattice to play a role,
so the overall picture is quite consistent with all data.
So, within the crude approximations in a toroidal compactification 
we have
\be 
\delta_H \sim   10^{-5} \quad \Longleftrightarrow \quad  
 M_s \sim M_{GUT}
\ee
This means the coupling unification in MSSM is consistent with
brane inflation in the braneworld. A more precise determination 
of the relation between $\delta_H$ and $M_s$ will be 
interesting for phenomenologically realistic compactifications.
Note that $\lp > \lpl$ and $M_s \theta \lpl  \ge 2 \pi$. 
If $\lp \gg \lpl$, $\dper$ is effectively 2. Otherwise,
we may have non-hypercubic compactification with $\dper=3$.
As we shall see in a moment, the qualitative properties of this case
will remain intact.
To get a crude estimate of the compactification sizes,
let $n_1+n_2 = 4 +1 =5$ for the standard model gauge group.
Crudely, with $\theta \simeq u/n_1$, we have $M_s u \rpl \sim 1$ and 
the 3 compactification radii
\be 
u \rpl : \rpl : \rp ~~\simeq ~~ 1 : u^{-1} : 400 g_s u
\ee
For $\theta \sim 0.1$, we have  $u \sim 1/3$, $M_s \lpl \sim 20$
and $M_s\lp \sim 10^3$.
In any case, $M_s \lesssim M_{GUT}$ is much closer to the $GUT$ scale 
than to the electroweak scale. 
As long as $M_s$ is close to $M_{GUT}$,
the power-running of the couplings from $\rpl^{-1}$ to $M_s$ will not 
mess up the coupling unification \cite{dienes}. In fact, it is even 
possible that the coupling unification may be improved slightly by a 
small power-running. A more careful analysis will be interesting. 

\subsection{Generic Compactification}

\DOUBLEFIGURE[t]{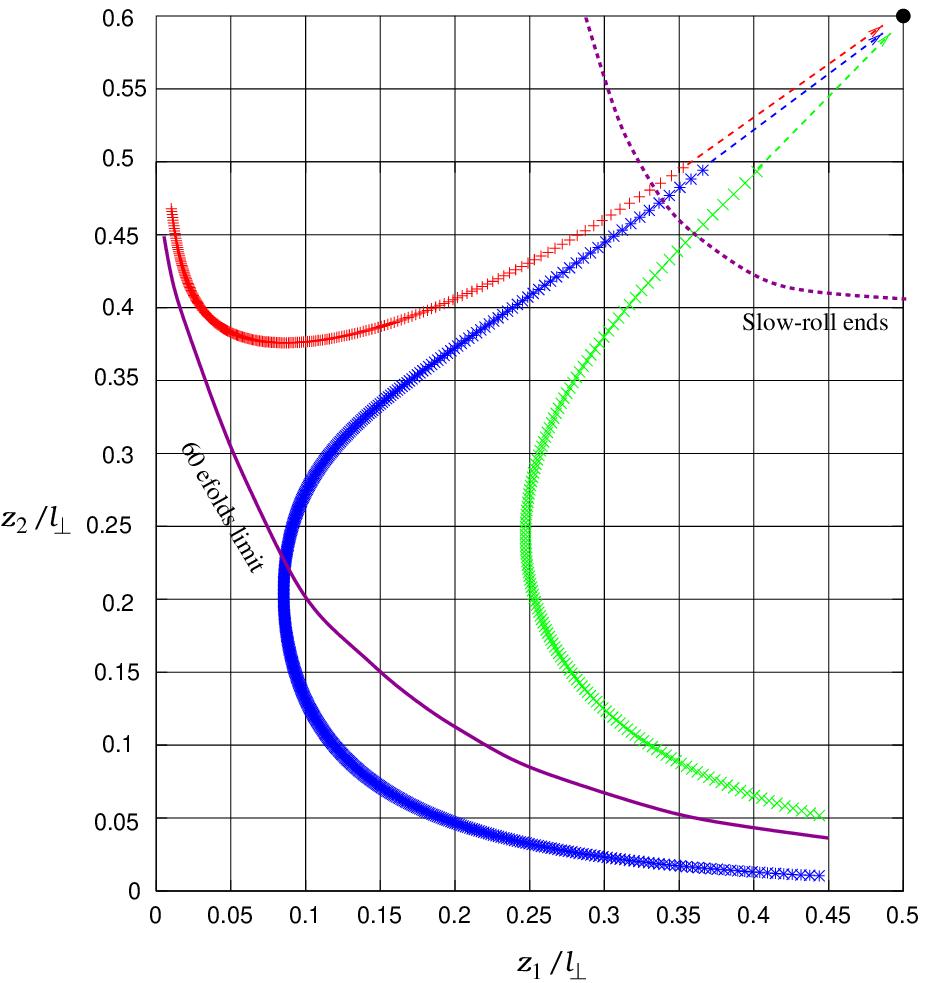,width=0.4\textwidth}
		{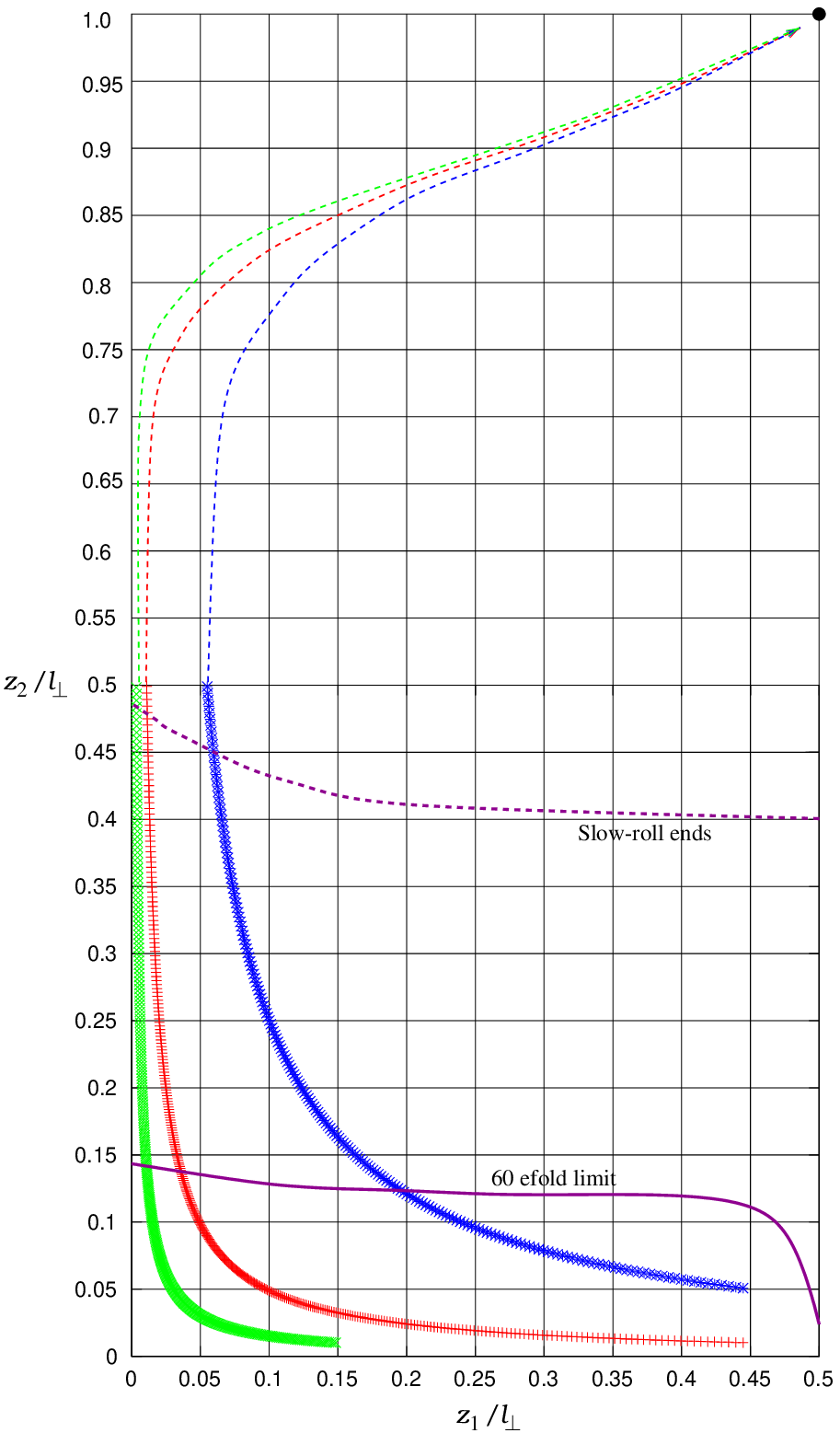,width=0.4\textwidth}
		{Rectangular lattice with small asymmetry, $a=1.2$. 
	One brane is sitting at the upper right corner. 
	There will be enough inflation if the other brane starts inside 
the 60 e-fold region. 3 possible trajectories of the inflaton are shown.}
		{Rectangular lattice with asymmetry $a=2$. 
3 possible trajectories of the inflaton are shown. The paths in the 
slow-roll region are reminiscent to renormalization group flows.} 

For hypercubic lattice, we see that it is easy to obtain a region that 
have enough inflation. Since the potential (\ref{Vhc}) has the form
$V(\varphi) - A \propto - z^4$, it is very flat around the antipodal 
point $z_i=0$, so it is easy to see that a region around $z_i=0$ will 
have enough inflation. For generic compactifications, lower powers of 
$z_i$ appears in the potential. As a result, the potential is much steeper,
so naively one may expect the valid region for enough inflation
to shrink rapidly to zero around the symmetric point where 
the net force is zero. However, the actual situation is more robust 
than naively exected. Here, let us illustrate the situation
with a rectangle-like lattice. As shown in Appendix, quadratic terms 
appear in the potential around the symmetric point. Let us consider
a 2-dimensional rectangular lattice, with sides $\lpl$ and $a\lpl$.
In this scenario, the effective potential becomes
\be
\label{Vrc}
V(\varphi) \simeq A \left(1 - \hat \lambda [\hat U (\varphi_2^2-\varphi_1^2)
+D\varphi_1^2\varphi_2^2 - C(\varphi_1^4 +\varphi_2^4) + \cdots]\right)
\ee           
where $U$ is a function of the ratio $a$ of the two sides of the 
rectangular lattice. Clearly $\hat U \rightarrow 0$ as $a \rightarrow 1$.
In Table 2, we show the values $U= \hat U /(\tau_p \Vpl \lp^2/2)$, 
$C$ and $D$ for $\dper=2$.

\begin{center}
\begin{tabular}{|c|c|c|c|}
\hline
a & U & C & D \\
\hline
\hline
1.0 & 0 & 3.94 & 23.64 \\
\hline
1.2 & 0.59 & 2.48 & 14.93 \\
\hline
1.5 & 0.97 & 1.08 & 6.48 \\
\hline
2.0 & 1.03 & 0.23 & 1.40 \\
\hline
\end{tabular}
\end{center}


The inflaton is a two-component field, so the paths for the inflaton 
can be quite complicated. Typical paths are shown in the figures.
In Figure 7 (for $a=1.2$) and Figure 8 (for $a=2$), one brane is
sitting at the upper right corner. If the other brane (at angle 
$\theta \sim 1/6$ with respect to the first one) starts inside the 
60 e-fold region around the lower left corner, there will be enough 
inflation. Comparing Figures 7 and 8 to Figure 6, each with the same 
angle $\theta \sim 1/6$, we see that ${\it P}$ is mildly sensitive 
to the shape (complex structure) of the torus 
\be
{\it P(\theta,a)} \simeq \frac{{\it P(\theta)}}{a}
\ee
for $a \ge 1$. However, the shape of the 60 e-fold
region and the inflaton path do depend strongly on the properties of 
the torus. In rectangular lattices, the quadratic $z$ (i.e., $\varphi$)
terms in the potential dominate 
during the initial part of the inflaton path. If the asymmetry is large,
the quadratic term is dominant over a large region, and the inflaton 
trajectories are simple, as can be seen in Figure 8.  
When the quartic terms become important they will cause the trajectories
to curve towards the corners of the cell. The location of the point 
where the trajectories turn depends on the asymmetry of the cell. 
Towards the end of inflation, when the branes are relatively close, 
the images become negligible and the potential is Coulombic.
When the brane separation is comparable to some of 
the compactified directions, the effects of the Kaluza-Klein modes 
should be included. It will be interesting to study their effect, 
although we believe the qualitative features discussed above should 
remain intact. 

By now, it is intuitively clear that the particular hypercubic 
compactification is not needed for a reasonable value of ${\it P}$. 
Generically, the images of one 
brane exert forces on the other brane, and there is always a point 
where the forces balance each other so the inflaton potential is flat 
at that point. Due to the softness of the brane interaction, there 
is always a region around that point where the potential is flat 
enough for sufficient inflation. For initially randomly placed branes 
at relatively small angle $\theta \lesssim 1/{10}$, which is not hard 
to arrange for generic compactifications, the probability of sufficient 
inflation is of order unity. This is very important, since,
in any realistic compactification in string model-building, a hypercubic 
lattice is a very special case. 

\subsection{Reheating and Defect Production}

We see that the inflaton and the tachyon modes are charged 
under the gauge symmetry. 
In more realistic models, there are chiral fields among the brane modes,
which certainly couple to the gauge fields.
During reheating after inflation, we expect the inflaton and the tachyon
modes to decay to the gauge bosons and the chiral fields. 
Since the inflaton and the tachyons are brane modes, we expect this 
reheating process to be very efficient. This picture is essentially 
that of hybrid inflation.

When $p$-branes collide, tachyons appear and energy is released.
Tachyon condensation generically allows the formation of 
lower-dimensional branes. Following Eq.(\ref{S2},\ref{SVall}),
\be
V_T \propto tr(TT^\dagger) \left( M_s^4 y^2 -
2 \pi M_s^2 \theta \right)
\ee
so when $y$ is small, the open string ground state $T$ becomes tachyonic
and the Higgs mechanism (\ref{higgs}) takes place. Here, the vacuum structure 
has a $U(N) \times U(N) /U(N)$ symmetry. Generically,
lower-dimensional D$(p-2k)$-branes can be formed \cite{k-theory} during the 
brane collision. Since this happens after inflation, it is important 
to check if the old monopole problem returns or not.
A naive application of the Kibble mechanism will 
be disastrous, since that will yield a defect per Hubble volume,
resulting in over-abundance of domain walls and monopole-like objects.
Fortunately this does not happen, due to the properties of the brane theory.

Imagine a string model that describes our world today. The early universe
of this model generically contains branes of all types. The 
higher-dimensional branes collide to produce lower-dimensional branes and 
branes that are present today. To be specific, consider a Type IIB 
orientifold, with D5-branes and orientifold planes.
In early universe, there are generically additional D9, D7, D5, D3 
and D1-branes with a variety of orientations. 
(Branes with even $p$ are non-BPS and decay rapidly.) 
Since the total R-R charges in the compactified directions 
must be zero, the branes must appear in sets with zero R-R charges, except for
D5-branes, which must appear so the total R-R charge between D5-branes 
and orientifold planes is zero. For example, 
if there are only two 9-branes, they must form a pair of D9-anti-D9-branes. 
Since D9-branes fill the 9-dimensional space, we expect 
the D9-anti-D9-brane pair to annihilate first. The resulting tachyon 
condensation generically produces lower-odd-dimensional branes. 
Brane on top of each other collides rapidly (unless they are BPS with 
respect to each other), leaving behind branes 
separated in the compactified directions. 

Consider the last two branes with 3 uncompactified spatial dimensions that 
are not in today's string ground state. As they approach each other, the 
universe is in the inflationary epoch. Suppose there are other branes 
with zero, one or two uncompactified dimensions, which appear as 
point-like objects, cosmic strings or domain walls, respectively. 
They are defects in 4-D spacetime. They either annihilate or are 
inflated away, resulting in negligible densities.
After inflation ends, the collision of the two branes with 3 
uncompactified spatial dimensions may produce lower-dimensional branes, 
which appear as defects. 
A large density of such defects may destroy the nucleosynthesis or even
overclose the universe (like the old monopole problem). 
There are two mechanisms to produce such defects: the Kibble mechanism and
thermal production. 

After inflation, when the ground state open string mode becomes tachyonic,   
brane collision and tachyon condensation takes place, and lower-dimensional 
branes will generically appear via the Kibble mechanism.  
To be specific, consider a D$p$-anti-D$p$-brane collision, which
may result in D$(p-2)$-branes and anti-D$(p-2)$-branes.
At this time, the particle horizon size is typically bigger than
the compactification sizes,
\be
\frac{1}{H} \simeq \frac{M_P}{M^2_s} \frac{(2 \pi)^{3/2}}{\theta} >> \lp
\ee
so the Kibble mechanism does not happen in the 
compactified directions. In the uncompactified directions, the Kibble 
mechanism can take place, so cosmic string-like defects may be formed:
they are D$(p-2)$-branes wrapping the same compactified cycles as the 
original $p$-branes, with one uncompactified dimension. 
Similarly, if the D$p$-brane collision can produce D$(p-4)$-branes, 
their production will be suppressed since there is less than one 
Hubble volume in the compactified directions. 

This implies that domain walls and monopole-like objects are not 
produced by the Kibble mechanism, while cosmic strings may.
Generically, there may be closed and stretched cosmic strings, and 
they form some sort of a network.
Fortunately, in contrast to domain walls and monopole-like objects,
a cosmic string network may be acceptable \cite{kolb}.
A more careful analysis of the production of cosmic strings will 
be very interesting.
This result is somewhat different from that of Ref.\cite{majumdar}, 
which does not take into account the compactification effect.

Next, let us consider thermal productions. We shall argue that thermal 
production is negligible. Branes can wrap around the 
compactification cycles. 
The mass of such a brane wrapping a p-dimensional volume $V_p$, 
which appears as a point-like object in 4 dimensional spacetime, is
\be
M_p \approx \tau_p V_p = \frac{M_s v_p}{2 \pi g_s} \ge \frac{M_s}{g_s} 
\ee 
where $v_p = M^p_sV_p/(2 \pi)^p \ge 1$. Since a wrapped brane is 
charged, while the total charge of the defects must be zero,
they have to be pair-produced or multi-produced. 
Assuming efficient reheating after inflation, the reheat 
temperature is given by 
\be
\label{temprh}
T_{RH} \approx \left(\frac{30 (2 \tau_p \Vpl)}{\pi^2 n_{dof}}\right)^{1/4}
\ee 
where $n_{dof}$ is the number of light degrees of freedom at reheating.
Typically $n_{dof}$ is of the order of a few hundred.
(For less efficient reheating, the temperature is of course lower.)
$T_{RH} \approx M_s/3$ for brane-anti-brane annihilation.
Comparing to $M_p$, this temperature is probably low enough to 
prevent the production of defects after inflation. 
This problem will be totally absent in the collision of branes  
at angle since the energy released is lower than that from the
brane-anti-brane annihilation. Also, if all the compactification 
directions are larger than the string scale (by a factor of 2 
is enough), the defects will be too heavy to be produced.
So the defect production is not a problem in this scenario.

\section{Discussions}

Although we find that the string scale is quite close to 
the $GUT$ scale, this does not imply that a low string scale (say 
$M_s=$ 10 TeV) is necessarily ruled out. Such a low string scale is 
possible if the radion grows a lot after inflation \cite{ira,radion}.
In this scenario, the model necessarily becomes rather complicated.

Keeping the radion mode as a dynamical variable,
we can integrate out the $d$ compactified dimensions
to obtain the low energy effective 4-dimensional action
\baray
S &=& \int d^4 x \sqrt{\det g} \, \bigg[ { e^{d {\tilde \Phi}}
\over 16 \pi G_N} {}^{(4)}R + { d (d-1) \over 16 \pi G_N } e^{d {\tilde
\Phi}} (\nabla {\tilde \Phi})^2
- V(\psi,{\tilde \Phi}) +
{\cal L}_{\rm sm}(g_{\mu\nu}, \psi) \bigg]
\label{action_Jordan}
\earay
The action (\ref{action_Jordan}) has the form of a scalar-tensor
theory of gravity, written in the Jordan frame.  
The Jordan-frame potential $V$ is given by
\be
V(\psi,{\tilde \Phi}) = r_0^d e^{d {\tilde \Phi}} V_{\rm bulk}
({\tilde \Phi}) - { k_i \over 8 \pi G_N r_0^2 } \, e^{(d-2)
{\tilde \Phi}}
+ V_{\rm brane}(\psi,{\tilde \Phi}),
\label{potentialJ}
\ee
where the Ricci scalar of the metric $h_{ab}$ is $2 k_i r_0^{-2}$
($r_0$ being the equilibrium radius of the extra dimensions today)
and $k_i$ is a dimensionless constant that we may neglect.
The action may be written in the Einstein frame,
\be
S = \int d^4 x \sqrt{- {\hat g}} \bigg[ {{\hat R} \over 16 \pi
G } - {1 \over 2} ({\hat \nabla} \Phi)^2
- {1 \over 2} e^{-\Phi/\mu} ({\hat \nabla} \psi)^2 -
e^{-2 \Phi/\mu}\, V(\psi, \Phi) \bigg]
+ S_{\rm rest}[e^{-\Phi/\mu} {\hat g}_{\alpha\beta},
\chi_{\rm rest}]
\label{action2}
\ee
where ${\hat g}_{\alpha\beta}$ is the Einstein frame metric,
and $g_{\alpha\beta} = e^{-\Phi/\mu} {\hat g}_{\alpha\beta}$ is
the physical, Jordan frame metric. Here,
$\mu = M_P \,  \sqrt{ {d+2 \over 8 d} }$, and
the canonically normalized radion field $\Phi$ is related to the 
radius $r$ of the extra dimensions by
$r = r_0 \exp [{\Phi \over  d \mu }]$.
The field $\psi$ is a brane scalar mode.
The quantity $V(\psi,\Phi)$ is the Jordan-frame potential for the
radion and inflaton (energy per unit proper brane 4-volume).
Finally the action $S_{\rm rest}[g_{\alpha\beta},\chi_{\rm rest}]$ is the
action of the remaining matter fields $\chi_{\rm rest}$, which may be
treated as a fluid.

For generic $V(\Phi)$, the factor $e^{-2 \Phi/\mu}$ will render the
potential $e^{-2 \Phi/\mu}  V(\Phi)$ unsuitable for inflation. 
(Such an exponential form is
too steep for slow-roll and too shallow for reheating.)
However, $\Phi$ may still play a role in an inflationary universe.
An example of this scenario has been considered in Ref.\cite{ira}.
With appropriate $V(\psi,\Phi)$, let $\psi$ be the inflaton, and 
$\Phi$ be frozen during the inflationary epoch. After inflation, $\Phi$
grows, resulting in today's value for the Newton's constant, which is
substantially smaller than the effective Newton's constant during 
inflation. This allows the correct value for density perturbation 
even though the string scale can be much smaller than what we find 
earlier. Of course, this scenario is much more complicated and 
model-dependent than that discussed in this paper.

\section{Summary and Remarks}

In brane inflation, the inter-brane separations play the role of 
inflatons, while the bulk modes provide the brane interactions that 
generate the inflaton potential. Such properties are well studied 
in string theory. By itself, the resulting potential is generically
unsuitable for inflation. 
Fortunately, any realistic phenomenological string 
realization requires the compactification of the extra dimensions.
Also, the weak coupling behavior observed in nature probably requires
the branes to have extra dimensions (beyond the 3 uncompactified 
spatial dimensions) that wrap around some of the compactification 
directions. These two properties dramatically improves the brane 
inflationary scenario. Without any fine-tuning, the probability 
of randomly placed branes in the early universe to originate 
substantial inflation can easily be of order unity.
This allows one to argue that brane interaction provides an 
explanation of the origin of inflation.

In a simple brane inflationary scenario suggested by our analysis, we 
find that the amplitude of the density perturbation observed by COBE 
implies that the string scale is very close to the $GUT$ scale. A more 
careful analysis in more realistic string models (e.g., orientifolds) 
can make this relation more precise. We can reverse the analysis: 
starting with MSSM and coupling unification, with $M_s=M_{GUT}$,
we find that a generic orientifold model will yield a density perturbation
in CMB of the correct magnitude.

A number of other issues also deserve further analysis, e.g., the 
likely braneworld just before inflation, the reheating and the defect 
production after inflation, $p$-brane-$p\prime$-brane interactions, the 
generation of density perturbation due to multi-component feature of the
inflaton, as is likely to be the case in the braneworld. We hope that 
this work shows that further studies along this direction is worthwhile. 
                                                             
We thank Keith Dienes, Sash Sarangi, Ashoke Sen, Gary Shiu and Ira 
Wasserman for discussions.
This research was partially supported by the National Science Foundation.  

\appendix
\section{Summing Over The Lattice}

Here we provide some details to the derivation of the potential
via summing over the lattice.
The potential for the D-dimensional lattice is given by:
\be
V\left(\overrightarrow r \right) = \left\{ \begin{array}{l}
A - B\sum_{i}\frac{1}{\left|\overrightarrow r -\overrightarrow r_i 
\right|^{\dper - 2}}, \dper > 2 \\
A + B\sum_{i}\ln\left|\overrightarrow r -\overrightarrow r_i \right|, 
\dper = 2
\end{array}
\right.
\ee
where the sum is over all the lattice sites. In order to estimate the 
potential around the antipodal point 
we expand to 4th order in $\ \overrightarrow z$, around the center of 
the elementary cell.
\be
\sum_{i}\frac1{\left|\overrightarrow z - 
\overrightarrow r_i\right|^{\dper}} = 
\sum_{i}\frac1{r_i^{\dper}\left[1 
- 2\overrightarrow z\cdot\overrightarrow r_i/r_i^2
 + z^2/r_i^2 \right]^{\dper/2}}
\ee
denoting $ \delta = -2\overrightarrow z \cdot \overrightarrow r_i/r_i^2 
+ z^2/r_i^2 $ and $p=\dper-2>0$ we have:
\baray
&& \sum_{i}\frac1{\left|\overrightarrow z - 
\overrightarrow r_i\right|^{p}} =
\sum_{i}\left[\frac1{\left(r_i^2\right)^{p/2}} - \frac{p}2\frac{\delta}
{\left(r_i^2\right)^{p/2+1}}  +
\frac{p}2\left(\frac{p}2+1\right)\frac1{2!}\frac{\delta^2}
{\left(r_i^2\right)^{p/2+2}} \nonumber \right. \\
&& - \frac{p}2\left(\frac{p}2+1\right)\left(\frac{p}2+2\right)
\frac1{3!}
\frac{\delta^3}{\left(r_i^2\right)^{p/2+3}} 
\nonumber \\
&& \left. + \frac{p}2\left(\frac{p}2+1\right)\left(
\frac{p}2+2\right)
\left(\frac{p}2+3\right)
\frac1{4!}\frac{\delta^4}{\left(r_i^2\right)^{p/2+4}} + \cdots
 \right]
\earay
For $p=0$ ($\dper=2$), the factor of $p$ in every term should be dropped.
Using the expression for $ \delta $ and grouping the 4th powers of 
$z$ we obtain:
\baray
&& V_4\left(\overrightarrow z\right) = 
\frac1{4!}\left[-p\left(p+2\right)
\sum_{i}\frac{3z^4}{r_i^{p+4}} +
p\left(p+2\right)\left(p+4\right)\sum_{i}\frac{6z^2
\left(\overrightarrow z\cdot\overrightarrow r_i\right)^2}{r_i^{p+6}}\right.
\nonumber \\
&& \left. -p\left(p+2\right)\left(p+4\right)
\left(p+6\right)\sum_{i}
\frac{\left(\overrightarrow z\cdot\overrightarrow r_i\right)^4}
{r_i^{p+8}}\right]
\earay
As an example, for a two-dimensional square lattice, we express 
$\overrightarrow z$ in terms of the components,  
$\overrightarrow z = (z_1, z_2)$ and obtain:
\baray
&& V_4\left(z_1, z_2\right) = \frac{1}{4!}\left(z_1^4+z_2^4\right)
\left[\sum_{i,j}\frac{48\left(i+\frac12\right)^2
\left(j+\frac12\right)^2}{\left(\left(i+\frac12\right)^2
+\left(j+\frac12\right)^2\right)^4} - 
\sum_{i,j}\frac{6}{\left(\left(i+\frac12\right)^2
+\left(j+\frac12\right)^2\right)^2} \right] \nonumber \\
&& -\frac{1}{4!}z_1^2z_2^2\left[\sum_{i,j}
\frac{288\left(i+\frac12\right)^2\left(j+\frac12\right)^2}
{\left(\left(i+\frac12\right)^2+\left(j+\frac12\right)^2\right)^4} - 
\sum_{i,j}\frac{36}{\left(\left(i+\frac12\right)^2
+\left(j+\frac12\right)^2\right)^2} 
\right]
\earay
Similarly, in  the case of a four dimensional hypercubic lattice,
$\overrightarrow z = (z_1, z_2, z_3, z_4)$, and the potential is:
\baray
&& V_4\left(z_1,z_2,z_3,z_4\right) = 
\frac1{4!}\left(z_1^4+\cdots+z_4^4\right)\left[192\sum_{i,j,k,l}
\frac{\left(i+\frac12\right)^2\left(j+\frac12\right)^2 + 
\cdots +\left(k+\frac12\right)^2\left(l+\frac12\right)^2 }
{\left(\left(i+\frac12\right)^2 + 
\cdots +\left(l+\frac12\right)^2\right)^5} \right. \nonumber \\
&& -\left. \sum_{i,j,k,l}\frac{48}{\left(\left(i+\frac12\right)^2+\cdots
+\left(l+\frac12\right)^2\right)^3} \right] 
\nonumber \\
&& -\frac1{4!}\left(z_1^2z_2^2+\cdots+z_3^2z_4^2\right)\left[384\sum_{i,j,k,l}
\frac{\left(i+\frac12\right)^2\left(j+\frac12\right)^2+
\cdots+\left(k+\frac12\right)^2\left(l+\frac12\right)^2}
{\left(\left(i+\frac12\right)^2+\cdots+
\left(l+\frac12\right)^2\right)^5} \right. \nonumber \\
&& - \left. \sum_{i,j,k,l}\frac{96}{\left(\left(i+\frac12\right)^2+\cdots
+\left(l+\frac12\right)^2\right)^3} \right]
\earay
If the lattice is not hypercubic, the 
leading term in the expansion is the 2nd order term.
Using a two-dimensional rectangular lattice as example, we write the 
vectors of the lattice as $\overrightarrow r_{i,j}=
\left(i+1/2\right)\hat{z_1}+a\left(j+1/2\right)\hat{z_2}$, where 
$\hat{z_1}$ and $\hat{z_2}$ are the unit vectors in the 
$z_1$ and $z_2$ directions and $a$ is the ratio of the two sides of 
the rectangular lattice. The second-order term of 
the expansion of the potential is:
\be
V_2\left(z_1,z_2\right) = \left(z_1^2-z_2^2\right)\frac{1}{2!}\sum_{i,j}
\frac{a^2\left(j+\frac12\right)^2-
\left(i+\frac12\right)^2}{\left(i+\frac12\right)^2 
+ a^2\left(j+\frac12\right)^2} 
\ee
We see that $V_2\left(z_1,z_2\right)$ becomes zero if $a = 1$.

We note that summing over the lattice will generically induce a 
correction to the constant $A$ in the potential. However, this 
correction is typically quite small and will be ignored.

\end{document}